\documentclass[pdftex,twocolumn,epjc3]{svjour3}          

\RequirePackage[T1]{fontenc}

\smartqed  

\usepackage{amsmath}
\usepackage{amsfonts}
\usepackage{amssymb}
\usepackage{graphicx}
\usepackage{mathptmx}
\usepackage{caption}
\usepackage{subfigure}
\usepackage{footnote}
\usepackage{makeidx}
\usepackage{color}

\RequirePackage{graphicx}
\RequirePackage{mathptmx}      
\RequirePackage{flushend}
\RequirePackage[numbers,sort&compress]{natbib}
\RequirePackage[colorlinks,citecolor=blue,urlcolor=blue,linkcolor=blue]{hyperref}

\journalname{Eur. Phys. J. C}

\begin{document}
    
    \title{Cosmological dynamics of the general non-canonical scalar field models}
    
    \author{ Jibitesh Dutta\thanksref{e1,addr1,addr2}
        \and
        Wompherdeiki Khyllep\thanksref{e2,addr3,addr4} 
        \and
        Hmar Zonunmawia\thanksref{e3,addr3}
    }
    
    \thankstext{e1}{e-mail: jdutta29@gmail.com, jibitesh@nehu.ac.in}
    \thankstext{e2}{e-mail: sjwomkhyllep@gmail.com}
    \thankstext{e3}{e-mail: zonunmawiah@gmail.com}
    
    \institute{ Mathematics Division, Department of Basic Sciences and Social Sciences, North Eastern Hill University, Shillong, Meghalaya 793022, India \label{addr1}
        \and
        Inter University Centre for Astronomy and Astrophysics, Pune 411 007, India \label{addr2}
        \and
        Department of Mathematics, North Eastern Hill University,  Shillong, Meghalaya 793022, India \label{addr3}
        \and
        Department of Mathematics, St. Anthony's College, Shillong, Meghalaya 793001, India\label{addr4}
    }
    
    \date{Received: date / Accepted: date}

    \maketitle
    
    \begin{abstract}
        We extend the investigation of cosmological dynamics of the general non-canonical scalar field models by dynamical system techniques for a broad class of potentials and coupling functions. In other words, we do not restrict the analysis to exponential or power-law potentials and coupling functions. This type of investigation helps in understanding the general properties of a class of cosmological models. In order to better understand the phase space of the models, we investigate the various special cases and discuss the stability and viability issues. Performing a detailed stability analysis, we show that it is possible to describe the cosmic history of the universe at the background level namely the early radiation dominated era, intermediate matter dominated era and the late time dark energy domination. Moreover, we find that we can identify a broad class of potentials and coupling functions for which it is possible to get an appealing unified description of dark matter and dark energy. The results obtained here, therefore, enlarge the previous analyses wherein only a specific potential and coupling functions describes the unification of dark sectors. Further, we also observe that a specific scenario can also possibly explain the phenomenon of slow-roll inflationary exit.    
        
    \end{abstract}
    
    \section{Introduction}
    The source for the observed accelerated expansion of the Universe still remains to be an obscured problem in modern cosmology. It is widely believed that this phenomenon can be theoretically explained by an exotic quantity commonly known as dark energy (DE) with large negative pressure \cite{Riess:1998cb,Perlmutter:1998np}.  In the literature, there are various candidates of DE and the debate over the fittest candidate of DE, is yet to be resolved. The details and status of various candidates of DE so far are summarised in an excellent review by Brax \cite{Brax:2017idh} recently. Scalar field plays a significant role in cosmology as they are sufficiently complicated to produce the observed dynamics. They are used to model DE and characterise inflation \cite{Liddle:2000cg}. Motivated by high energy physics, scalar field models play an important role to explain the nature of DE due to its simple dynamics \cite{Copeland:2006wr,Tsujikawa:2013fta}. The simplest form of a scalar field is the canonical field also known as quintessence field with potential. However, the canonical scalar field cannot fully explain several complex cosmological dynamics of the Universe. For instance, the quintessence model cannot explain the crossing of the phantom divide line, bouncing solutions. This leads to a more general description of a scalar field known as non-canonical scalar field \cite{Fang:2006yh,Mukhanov:2005bu}. Another advantage of the non-canonical setting over canonical is that it can resolve the coincidence problem without any fine-tuning issues \cite{Lee:2014bwa}.

    The most general non-canonical form of a scalar field which involves higher order derivatives of a scalar field falls under the well-known Horndeski Lagrangian \cite{Horndeski:1974wa}. Within this class, a simple form of a non-canonical scalar field model is collectively known as $k$-essence, the first term of the Horndeski Lagrangian. Basically, $k$-essence models are the generalisation of the canonical scalar field and its Lagrangian is a function of both the scalar field and its kinetic term \cite{ArmendarizPicon:2000ah}. An important feature of the $k$-essence is that, unlike the canonical scalar field, its kinetic energy term can also source acceleration. In the context of cosmology, the $k$-essence field was first used to describe early inflationary epoch in \cite{ArmendarizPicon:1999rj} and was later found that it can also describe DE by Chiba {\it et al.} \cite{Chiba:1999ka}. In literature, there are several forms of  $k$-essence models depending on the type of its Lagrangian \cite{Mukhanov:2005bu,ArmendarizPicon:2005nz,Unnikrishnan:2008ki}.  
    
   In the present work, we will consider a general form of $k$-essence Lagrangian introduced by Melchiorri {\it et al.} \cite{Melchiorri:2002ux}. It has been found that while the potential energy of this type of  non-canonical scalar field models behaves as DE, its kinetic energy plays the role of dark matter (DM) \cite{Kamenshchik:2001cp,Padmanabhan:2002cp,Scherrer:2004au,DiezTejedor:2006qh,DeSantiago:2011qb,Sahni:2015hbf,Mishra:2018tki}. This leads to a unification of DE and DM, allowing structure formation and avoiding the strong integrated Sachs-Wolfe effect in the CMB anisotropies which usually afflict the unification between dark sectors \cite{Bertacca:2008uf}. Interestingly, the Chaplygin gas and its generalised form can also describe the DE-DM unification \cite{Kamenshchik:2001cp,bil,ben}. But the Chaplygin gas is associated with gravitational clustering problems and therefore cannot describe the real universe \cite{bar,sen}. Another advantage of the non-canonical models is that the adiabatic sound speed is sensitive to the values of a non-canonical parameter which have an effect on the gravitational clustering \cite{Sahni:2015hbf}. Moreover, cosmological observations such as SNIa, CMB and BAO favour  the general $k$-essence models\cite{Mamon:2015qai}.
    
    Dynamical system techniques are useful tools to investigate the complete asymptotic behaviour of a given cosmological model which allows us to bypass the difficulty in solving non-linear cosmological equations.  These tools also provide a description of the global dynamics of the universe by analysing the local asymptotic behaviour of critical points of the system and relate them with the main cosmological epochs of the universe. For instance, while a late-time DE domination would typically correspond to a stable point, the radiation and matter dominated eras correspond to saddle points. Dynamical system techniques had been extensively used in literature to analyse the evolution of various cosmological models of DE as well as modified gravity based models \cite{Copeland:1997et,Carloni:2007eu,Leon:2012mt,Leon:2013qh,Carloni:2015lsa,Boehmer:2015kta,Boehmer:2015sha,Dutta:2017fjw,Dutta:2017kch,Dutta:2017wfd,Dutta:2016bbs,Zonunmawia:2017ofc,Hohmann:2017jao}. For a recent and comprehensive review on the applications of dynamical system tools to various cosmological models see reference \cite{Bahamonde:2017ize}. However, one limitation of the dynamical system approach is the dependence on the choice of variables which characterise the solution associated with critical points. It is worth mentioning that the absence of some cosmological epoch of the Universe does not always imply the inability of the theory to explain such epoch, but it may be due to the inability of the associated dynamical system to exhibit the presence of an epoch.  
    
    In the non-canonical scalar field settings, the dynamical system tools have been applied extensively to various subclasses of $k$-essence Lagrangian \cite{Bahamonde:2017ize}. A dynamical system  analysis for a $k$-essence Lagrangian introduced in  \cite{Melchiorri:2002ux} has been analysed in \cite{DeSantiago:2012nk} for a power-law and exponential potential. Further, dynamical system analysis of an interacting DE within the framework of the $k$-essence Lagrangian introduced in  \cite{Melchiorri:2002ux} with power-law potential has been performed in \cite{Das:2014yoa}. In \cite{DeSantiago:2012nk,Das:2014yoa}, the choice of potential is restricted only to power-law and exponential case due to the choice of dynamical variables considered. However, dynamical analysis of the general non-canonical scalar field beyond specific potentials have never been studied before. On the other hand, there are various studies of the canonical scalar field beyond exponential potential \cite{syc,fang,mat,lop}. The important reason for studying the dynamical properties of DE model beyond a specific potential is that it helps in understanding general properties of a DE model which in turn helps in predicting properties of a class of DE models. Therefore, it is more economic and scientific to study for arbitrary potentials. Furthermore, the generalisation to various scalar field potentials is not only important from the mathematical point of view but also to connect these phenomenological models with the low energy limit of high energy physical theories. Therefore, it is interesting to extend the analysis to encompass the various class of potentials.
    
    With this motivation, here we extend the work of \cite{DeSantiago:2012nk} by performing a dynamical analysis for a more general $k$-essence Lagrangian over a wider class of scalar field potentials. Moreover, we shall consider a different choice of dynamical variables from the one defined in \cite{DeSantiago:2012nk} so as to encompass the desired class of potentials.  The main goal of this work is to determine the choice of model parameters which leads to interesting background cosmological dynamics such as the well-known unification of DE and DM within the non-canonical scenario. The present work might also provide a preliminary test of the general non-canonical models which are of interest for further investigation.

    The organisation of the paper is as follows: In Sect. \ref{non-canonical}, we briefly review the basic equations of the non-canonical scalar field models. In Sect. \ref{dynamical system}, we perform a dynamical system analysis for the autonomous system of differential equations obtained from the basic cosmological equations. Within this section, we also discuss the cosmological dynamics by considering a different class of scalar fields by taking various types of potentials and non-canonical coupling functions in various subsections. Finally, the conclusion is given in Sect. \ref{conclusion}.
    
    {\it Notation}: In this work, we shall assume the $(+,-,-, -)$ signature convention for the metric. We shall also adopt units where $8 \pi G = c = 1$.

    \section{The general non-canonical scalar field model and its basic cosmological equations}\label{non-canonical}
    The general action of a minimally coupled scalar field model is given by 
    \begin{equation}\label{action}
    S=\int d^4x\sqrt{-g}\Big(\frac{R}{2}+\mathcal{L}(\phi, X)\Big)+S_m,
    \end{equation}
    where  $\mathcal{L}(\phi, X)$ is the Lagrangian density which is an arbitrary function of a scalar field $\phi$ and its kinetic term $X$ ($X=\frac{1}{2} \partial_{\mu} \phi \partial^{\mu} \phi$), $R$ is the Ricci scalar and $g$ is the determinant of the metric $g_{\mu\nu}$. The last term $S_m$ represents the action of the background matter field.

    Varying the action (\ref{action}) with respect to the metric yields the Einstein field equations as
    \begin{equation}\label{einstein}
    G_{\mu \nu}=T^{\phi}_{\mu \nu}+T^m_{\mu \nu},
    \end{equation}
    where $G_{\mu\nu}$ is the Einstein tensor, $T^{\phi}_{\mu \nu}$ is the energy-momentum for $\phi$ given by
    \begin{equation}
    T^{\phi}_{\mu \nu}=\frac{\partial \mathcal{L}}{\partial X}\partial_{\mu}\phi \partial_{\nu}\phi-g_{\mu \nu}\mathcal{L}.
    \end{equation} 
    The energy-momentum tensor $T^m_{\mu \nu}$ for the matter
    component which is modeled as a perfect fluid is given by
    \begin{equation}
    T^m_{\mu \nu}=(\rho_m+p_m)u_\mu u_\nu+p_m\,g_{\mu \nu},
    \end{equation}
    where $\rho_m$ and $p_m$ are the energy density and pressure of the matter component and $u_\mu$ is the four velocity of the fluid.
    
    In the present work, we shall consider a homogeneous, isotropic and spatially
    flat Friedmann Robertson Walker (FRW) universe which is characterised by the line element
    \begin{equation}\label{eq9}
    ds^2=dt^2-a^2(t)[dr^2+r^2d\theta^2+r^2\sin^2\theta d\phi^2],
    \end{equation}
    and the following non-canonical Lagrangian density of the scalar field \cite{Melchiorri:2002ux}
    \begin{equation}\label{eq4}
    \mathcal{L}(\phi, X)=f(\phi)F(X)-V(\phi),
    \end{equation} 
    where $a(t)$ is the scale factor, $t$ is the cosmic time, $V(\phi)$ is a self-interacting potential for the scalar field $\phi$, $f(\phi)$ is an arbitrary function of $\phi$ with $f(\phi) \geq 0$ and $F(X)$ is an arbitrary coupling function of $X$. For a spatially homogeneous FRW metric \eqref{eq9}, $X=\frac{1}{2}\dot{\phi}^2$. The positive semi-definiteness of function $f$ is required to explain various cosmological observations as discussed in \cite{Melchiorri:2002ux}. It can be seen that Eq. (\ref{eq4}) reduces to quintessence field when $f(\phi)=1$ and $F(X)=X$, it reduces to phantom field when $f(\phi)=1$ and $F(X)=-X$. It means that each quintessence or phantom scalar field is equivalent to a particular $k$-essence model. This type of scalar field model \eqref{eq4} constitutes an alternative model of DE yielding late time accelerated solutions and it is well motivated from high-energy physics \cite{Tsujikawa:2004dp,Piazza:2004df}. 
    
    There are several functional forms of  $F(X)$ proposed so far in the literature see \cite{Mukhanov:2005bu,ArmendarizPicon:2005nz,Unnikrishnan:2008ki}. However, in order to obtain a concrete description of the cosmological dynamics, we shall consider a case where $F(X)=X^\alpha$ i.e. the non-canonical scalar field models whose Lagrangian density is given by  \cite{Mukhanov:2005bu,Unnikrishnan:2008ki} 
    \begin{equation}\label{eq6}
    \mathcal{L}(\phi, X)=f(\phi) X^\alpha-V(\phi),
    \end{equation}  
    where $\alpha$ is a dimensionless parameter. This class of $F(X)$ is a simple generalisation of the canonical scalar field (for  $\alpha=1$).   It leads to some rich phenomenology that is not present in the canonical case for e.g. the existence of non-singular bouncing solutions \cite{DeSantiago:2012nk}. This choice of $F(X)$ also allows a non-canonical field to cluster and depending on values of $\alpha$, it behaves either like warm or cold dark matter at small scales \cite{Sahni:2015hbf}. It has been found that the slow-roll conditions can be easily realized for the non-canonical scalar field models ($\alpha \neq 1$) compared to the corresponding canonical case ($\alpha=1$) \cite{Unnikrishnan:2012zu}. Further, non-canonical
models reduce the tensor-to-scalar ratio than their
canonical counterparts, leading to a better agreement with CMB observations \cite{Unnikrishnan:2012zu}. These interesting features of non-canonical scalar fields motivate us to further investigate the cosmological dynamics of such fields in a more general context.
    
    In terms of Lagrangian density $\mathcal{L}$, the expressions of the energy density $\rho_\phi$ and pressure $p_\phi$ associated with the scalar field are given by
    \begin{eqnarray} 
    \rho_\phi&=&2X\Big(\frac{\partial \mathcal{L}(\phi, X)}{\partial X}\Big)-\mathcal{L}(\phi, X)\,, \label{p_rho1}\\
    p_\phi&=&\mathcal{L}(\phi, X). \label{p_rho2}
    \end{eqnarray}
    Employing the line element \eqref{eq9}, the Einstein field equations can be written as
    \begin{eqnarray}
    3H^2&=&(2\alpha-1)f(\phi)\,X^{\alpha}+V(\phi)+\rho_m, \label{frd_eqn}\\
    \dot H&=&-\frac{1}{2}\Big[2\alpha f(\phi)\, X^{\alpha}+\rho_m(1+w)\Big], \label{ryc_eqn}
    \end{eqnarray}
    where the over-dot denotes differentiation with respect to cosmic time $t$, $H=\frac{\dot{a}}{a}$ is the Hubble parameter and $w$ is the equation of state (EoS) of matter  $(-1\leq w \leq 1)$ defined as $p_m=w \rho_m$. 
    
 Substituting for $\mathcal{L}$ from \eqref{eq6} into \eqref{p_rho1} and \eqref{p_rho2}, the energy density and pressure of the scalar field are respectively given by
    \begin{eqnarray}
    \rho_\phi&=&\rho_X+V,\label{eq7}\\
    p_\phi&=&p_X-V, \label{eq8}
    \end{eqnarray} 
    where $\rho_X=(2\alpha-1)\ f(\phi)\,X^{\alpha}$ ~\text{and}~ $p_X=f(\phi)\, X^{\alpha}$. 
    Finally, we assume that the scalar field and barotropic fluid energy momentum tensors are conserved separately i.e. there is no exchange of energy between the scalar field and barotropic fluid. In that case by employing line element \eqref{eq9}, the energy conservation equations are given by
    \begin{eqnarray}
    \dot\rho_\phi+3H(\rho_\phi+p_\phi)&=&0, \label{eq12}\\
    \dot{\rho}_m+3H\rho_m(1+w)&=&0.\label{eq13}
    \end{eqnarray}
    Using Eq. (\ref{eq7}) in Eq. (\ref{eq12}), the evolution equation of the scalar field can be expressed as
    \begin{eqnarray}\label{eq14}
    \ddot{\phi}&=&-\Big[\frac{1}{2\alpha}\frac{df}{d\phi}\frac{\dot{\phi}^2}{f(\phi)}+\frac{1}{\alpha(2\alpha-1)}\frac{dV}{d\phi}\Big(\frac{2}{\dot \phi^2}\Big)^{\alpha-1}\frac{1}{f(\phi)}\nonumber\\&&+\frac{3H\dot{\phi}}{2\alpha-1}\Big].
    \end{eqnarray}
    As expected, the above equation reduces to the case of a canonical field, $\ddot{\phi}+3H\dot \phi +\frac{dV}{d\phi}=0$, when $\alpha=1$ and $f(\phi)=1$.  It is worth noting that for constant potential, the kinetic part $\rho_X$ and the potential part $V(\phi)$ behave as two non-interacting fluids as they separately satisfy the conservation equation \eqref{eq12}. 

 For theoretical consistency of the Horndeski's most general scalar-tensor theories, the conditions $Q_s > 0$, $C_s^2 \geq 0$, $Q_t > 0$,  $C_t^2 \geq 0$ are required for the avoidance of ghosts and Laplacian instabilities associated with a scalar and tensor  perturbations \cite{DeFelice:2011bh}. In this model, we have verified  that the conditions  $Q_t> 0$,  $C_t^2 \geq 0$ holds for any choices of model parameters. However, the quantities $Q_s$ and $C_s^2$ are given by
    \begin{eqnarray}
    Q_s&=& 
  2^\alpha \alpha (2\alpha-1)H^{-2} X^{\alpha} f  \,\label{Qs}\\
    C_s^2&=&\frac{1}{2\alpha-1}+\frac{3(1+w)\rho_m}{2^\alpha \alpha (2\alpha-1)  X^{\alpha} f}\label{cs2}
    \end{eqnarray}
Therefore in the absence of barotropic matter component, the Eq. \eqref{cs2} reduces to  the case obtained in  \cite{Mishra:2018tki} given by
\begin{align}
C_s^2=\frac{1}{2\alpha-1}\,.
\end{align} 
It is important to  note here  that we have  $Q_s>0$ and $C_s^2 \geq 0 $ for $\alpha>\frac{1}{2}$. However,  for $\alpha=\frac{1}{2}$ or at a point where $X=0$ or $f=0$, the speed of sound $C_s^2$ diverges. This result is expected as discussed in Ref. \cite{Chimento:2004ha} for a typical form of $k$-essence Lagrangian. The divergence of speed of sound is a well known result in $k$-essence model and therefore violates causality \cite{Bonvin:2006vc}. Further the existence of superluminal solutions i.e. $C_s^2>1$  is a familiar characteristic of Horndeski model in the presence of external normal matter \cite{vik} and does not necessarily imply pathologies of a cosmological solution. Hence, to avoid a possibility of ghost instabilities ($C_s^2<0$), in what follows, we shall investigate  only for the case of  $\alpha>\frac{1}{2}$. In the next section, we shall convert these cosmological equations into an autonomous system of equations and perform a dynamical system analysis for various types of $f(\phi)$ and $V(\phi)$. 
    
    \section{Dynamical system analysis}\label{dynamical system}
    In order to write the above set of cosmological equations as an autonomous system of ordinary differential equations, we introduce the following set of normalised phase space variables

    \begin{eqnarray}\label{dmv}
    x=\frac{\dot \phi}{\sqrt{6}H}, \qquad y=\frac{\sqrt{V}}{\sqrt{3}H}, \qquad z=H^{2(\alpha-1)}f, \nonumber\\s=-\frac{1}{V}\frac{dV}{d \phi}, \qquad v=-\frac{1}{f}\frac{df}{d \phi}.~~~~~~~~~~~~~~~~~~~~~~~~~
    \end{eqnarray}
    Using these variables \eqref{dmv}, the above cosmological equations can be converted to the following autonomous system as
    
    \begin{eqnarray}
    x'&=&\frac{1}{\sqrt{6}}\Big[\frac{3x^2v}{\alpha}+\frac{3 y^2s}{z x^{2(\alpha-1)} \alpha(2\alpha-1)3^{(\alpha-1)}}-\frac{3\sqrt{6}x}{(2\alpha-1)}\Big] \nonumber\\&&  +\frac{x}{2}\Big[2\alpha3^\alpha x^{2\alpha}z  +3(1+w)\{1-3^{(\alpha-1)}x^{2\alpha}z(2\alpha-1)\nonumber\\&&-y^2\}\Big],~\label{xprime}\\
    y'&=&-\frac{xys}{2}+\frac{y}{2}\Big[2\alpha3^\alpha x^{2\alpha}z+3(1+w)\{1-3^{(\alpha-1)}x^{2\alpha}z(2\alpha-1)\nonumber\\&&  -y^2\}\Big],~\label{yprime}\\
    z'&=&-\sqrt{6}xzv-\frac{2z(\alpha-1)}{2}\Big[2\alpha3^\alpha\,  x^{2\alpha}z\nonumber\\&&+3(1+w)  \{1-3^{(\alpha-1)}x^{2\alpha} z(2\alpha-1)-y^2\}\Big],~\label{zprime}\\
    s'&=&-\sqrt{6}xg(s),~\label{sprime}\\
    v'&=&-\sqrt{6}xh(v),~\label{vprime}
    \end{eqnarray}
    where $g(s)=s^2(\Gamma_1-1)$ and $h(v)=v^2(\Gamma_2-1)$ with ~$\Gamma_1=V\frac{d^2V}{d \phi^2}\Big/ \Big(\frac{dV}{d \phi}\Big)^2$, $\Gamma_2=f\frac{d^2f}{d \phi^2}\Big/ \Big(\frac{df}{d \phi}\Big)^2$ and the prime denotes differentiation with respect to $N$= $\ln a$. Note that if $s=s(\phi)$ is invertible, then we can also express $\phi$ as function of $s$. Since the parameter $\Gamma_1$ is a function of $\phi$ only, hence $\Gamma_1$ can also be expressed  as $\Gamma_1=\Gamma_1(s)$. Similarly, we can write $\Gamma_2$ as function of $v$. This does not include any arbitrary $V$ or  $f$ but it includes a specific class of potential $V$ and coupling function $f$. Thus, for such choices of $V$ and $f$, the system \eqref{xprime}-\eqref{vprime} constitutes an autonomous system of equations. Here, we observed  that in the context of cosmological evolution, there are four basic variables $H$, $\phi$, $\dot{\phi}$ and $\rho_m$, which are related by a Friedmann constraint \eqref{frd_eqn}. Therefore, there are only three independent variables altogether. However, the introduction of two extra variables allows us to track the effect of the coupling and potential functions on the overall dynamics. In the literature, these types of variables are often introduced to study the class of unknown functions \cite{fang,leyva}.  It is worth noting that the variables \eqref{dmv} are different from those defined in \cite{DeSantiago:2012nk} and  this choice of variables helps in obtaining cosmologically viable critical points which however cannot be captured in \cite{DeSantiago:2012nk}. More importantly, our choice of variables leads to a viable cosmological sequence: radiation era $\rightarrow$ matter era $\rightarrow$ DE era and also can possibly explain a unified description of DM and DE. Moreover, the above system \eqref{xprime}-\eqref{vprime} reduces to the corresponding system of a canonical case for $\alpha=1$ and $f=1$ \cite{fang}. 
    
    It can be seen from the above system that $y=0$ is an invariant sub-manifold \footnote{The simplest approach to determine the existence of an invariant sub-manifold of $\mathbb{R}^n$ (here $n=5$) is to look at the form of the right-hand side (rhs) of a dynamical equation for a given variable (for e.g. $y$). If the dynamical equation is of the form $y'=(y-\xi) g$, where $g: \mathbb{R}^n \rightarrow \mathbb{R}$ is a continuous function, then $y=\xi$ is an invariant sub-manifold with respect to a flow corresponding to an autonomous vector field given by \eqref{xprime}-\eqref{vprime} \cite{tava}.}. Similarly, $z=0$ is an invariant sub-manifold, however, it is singular as $z$ appears as the denominator in Eq. \eqref{xprime}. For constant potential $V$, $x=0$ also represents an invariant sub-manifold of the system. Depending on the choice of $V$ and $f$, $s=0$ and $v=0$ may be invariant sub-manifolds. Invariant sub-manifolds play an important role in the characterisation of the phase space as these manifolds separate the phase space into independent sections. The presence of invariant sub-manifolds tells us that there is no global attractor except those stable points which lie on the intersection of all the invariant sub-manifolds \cite{Bahamonde:2017ize}.  The existence of accelerated global attractors guarantees late time evolution of the universe, irrespective of the choice of model parameters. Throughout the work, $s_*$ and $v_*$ denote the zeroes of $g(s)=0$ and $h(v)=0$ respectively. Further, $dg(s)$, $dh(v)$ denote the derivative of $g$ and $h$ with respect to $s$ and $v$ respectively. It is worth noting from the definition \eqref{dmv}, $y>0$ corresponds to $H>0$ i.e. expanding universe and $y<0$ corresponds to $H<0$ i.e. contracting universe. However, as the system \eqref{xprime}-\eqref{vprime} is invariant under the transformation $y \rightarrow -y$, we shall only analyse the case of expanding universe which is also cosmologically viable. Further, as we consider expanding universe, therefore we have $z \geq 0$.

  In terms of variables (\ref{dmv}), the relative energy density due to the scalar field $\Omega_\phi$, the relative energy density due to the barotropic matter $\Omega_{m}$ and the scalar field EoS $w_\phi$ can be expressed as
    \begin{eqnarray}
    \Omega_\phi&=&\frac{\rho_\phi}{3H^2}=(2\alpha-1)3^{(\alpha-1)}x^{2\alpha}z+y^2, \label{eq25}\\
     \Omega_{m}&=&\frac{\rho_m}{3H^2}=1-(2\alpha-1)3^{(\alpha-1)}x^{2\alpha}z-y^2,\label{eq26}\\
    w_{\phi}&=&\frac{p_\phi}{\rho_\phi}=\frac{3^{(\alpha-1)}z x^{2 \alpha}-y^2}{(2\alpha-1)3^{(\alpha-1)}z x^{2\alpha}+y^2}. \label{w_phi}
    \end{eqnarray}
    Note that the relative energy density parameters are related by the Friedmann constraint \eqref{frd_eqn}  as 
    \begin{equation}\label{frd_cons}
    \Omega_m+\Omega_\phi=1. 
    \end{equation}
    Here we see that for $\alpha>0$ if $x\approx 0$ and $y \approx 1$, then $w_\phi \approx -1$.   Therefore, when the potential energy of the scalar field dominates, the scalar field behaves as cosmological constant (i.e. DE).  However, when $y\, \approx \, 0$ and $(2\alpha-1) 3^{(\alpha-1)}x^{2\alpha}z \approx 1$ i.e. the contribution from the potential is negligible and maximum contribution comes from the kinetic part, we have $w_\phi \approx \frac{1}{2\alpha-1}$.  This implies that purely kinetic non-canonical models cannot describe the cosmic acceleration for $\alpha>\frac{1}{2}$ as reviewed in \cite{sari}. Further, we see that the scalar field  scales as radiation for $\alpha=2$, stiff matter for $\alpha=1$ and as dust for very large $\alpha$. This does not mean that the matter component comes from the DE, but both of them are sourced by a single scalar field. Additionally, we can define the effective energy density and pressure as
    \begin{eqnarray}
    \rho_{\rm eff}&=& \rho_{\phi}+\rho_m,\\
    p_{\rm eff}&=&p_{\phi}+p_m,
    \end{eqnarray}
    
    \noindent where we can obtain the effective EoS as
    \begin{eqnarray}\label{27}
    w_{\rm eff}&=&-1+\frac{2}{3}\,\alpha\,{3}^{\alpha}{x}^{2\,\alpha}z+ \left( w+1 \right) [ 1-{3}^{\alpha-1}{x}^{2\,\alpha}z \left( 2\,\alpha-1 \right)\nonumber\\&& -{y}^{2}], \label{eq28}
    \end{eqnarray}
    
    \noindent which is usually related to the deceleration parameter $q$ through
    \begin{eqnarray}\label{29}
    q=-1-\frac{\dot H}{H^2}=\frac{1+3w_{\rm eff}}{2}.
    \end{eqnarray}
    Hence for accelerated universe, one must obtain $q<0$ i.e. $w_{\rm eff}<-\frac{1}{3}$. From Eq. \eqref{frd_cons}, the physical requirement $\rho_m \geq 0$ yields the constraint
    \begin{equation}\label{eq23}
    \Omega_\phi \leq 1.
    \end{equation}
    
    \noindent Hence the five dimensional phase space of the system (\ref{xprime})-(\ref{vprime}) in terms of variables \eqref{dmv} is given by
    
    \begin{eqnarray}\label{eq24}
    \Psi&=&\left\lbrace (x,y,z) \in \mathbb{R}^3 : (2\alpha-1)3^{(\alpha-1)}x^{2 \alpha}\,z+y^2\leq 1, y, z \geq 0 \right\rbrace \nonumber\\&& \times \left\lbrace (s, v) \in \mathbb{R}^2\right\rbrace.
    \end{eqnarray}
   To determine the evolution of relevant energy densities numerically, we estimate the evolution of the Universe in such a way that it is consistent with the present observational data ($\Omega_{m} \simeq 0.3$, $w_{\rm eff}\simeq-0.7$) \cite{ade:2015xua}. This corresponds to points in the phase space where $x_0=0$, $y_0=\pm 0.83$, $z_0$ is any value.

    In order to extract the dynamics of the above system, first, we need to find the critical points of the system by equating the right-hand side of the system \eqref{xprime}-\eqref{vprime} to zero. This is followed by perturbing the system near the critical points from which the stability and type of critical points can be determined from the eigenvalues of the corresponding perturbed matrix. In the present work, since we have arbitrary functions $f$ and $V$, in what follows, we shall analyse various cases separately for different choices of $f$ and $V$.

    \subsection{$f(\phi)$=Constant, $V(\phi)=$Constant}\label{f_V_constant}
\begin{figure}
        \centering
        \includegraphics[width=6cm,height=5cm]{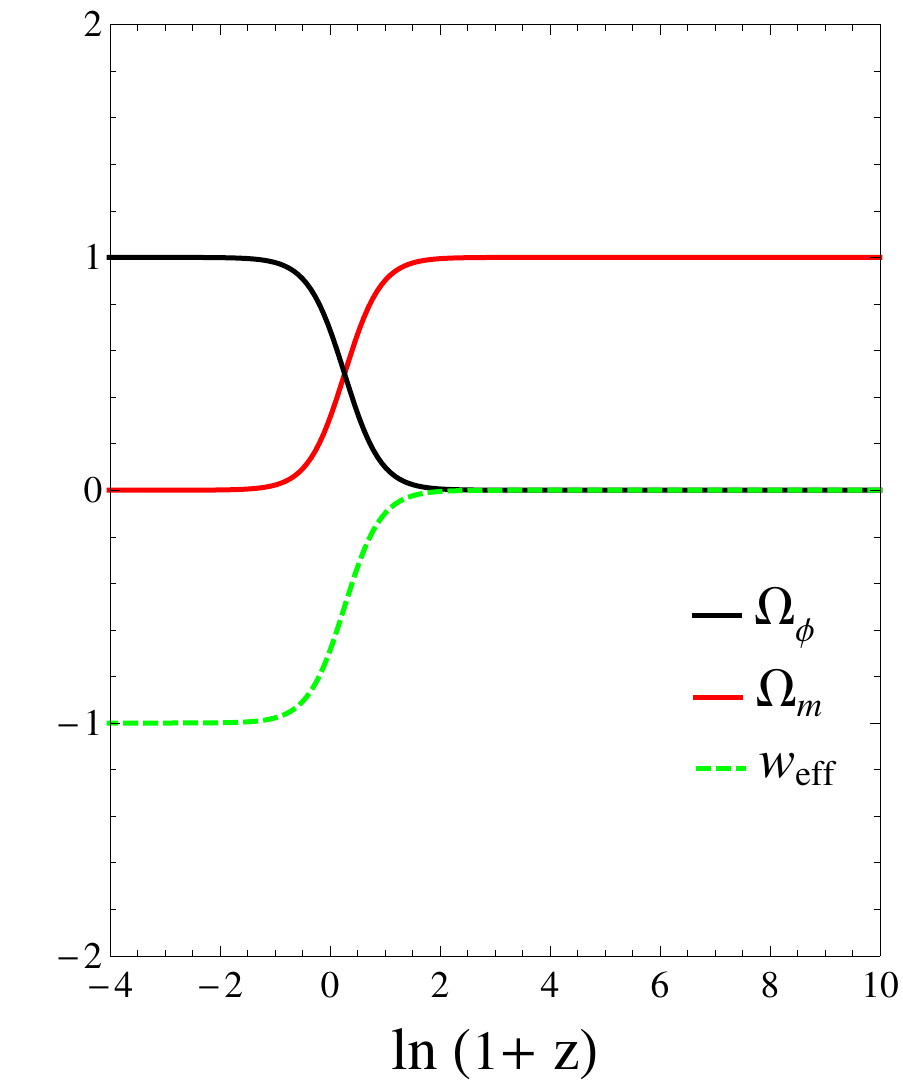}
        \caption{The time evolution of the relative matter energy density $\Omega_m$, the relative scalar field energy density $\Omega_\phi$ and the effective EoS $w_{\rm eff}$ for the case when $f=$ constant, $V=$constant. Here, $w=0$, $\alpha=2$.}
        \label{fig:Cos_Par1}
    \end{figure}

    We start our study by considering the simplest choice of $f$ and $V$ where both are constant. In this case the variables $s=0$ and $v=0$, hence, the system \eqref{xprime}-\eqref{vprime} reduces to a three dimensional system in $x$, $y$ and $z$. It is worth noting that for this choice of $V$ and $f$, the functions $\Gamma_1$, $\Gamma_2$ cannot be expressed as explicit functions of $s$ and $v$ respectively but the three dimensional system is still autonomous as the equations involving $\Gamma_1$, $\Gamma_2$ are not involved. We note here that the system has three invariant sub-manifolds $x=0, y=0$ and $z=0$. Under this case, we have only one critical point $A_1=(0, 0,0)$ and one set of critical points $A_2=(0, 1,z)$. 
    \begin{itemize}
        \item The critical point $A_1$ corresponds to a matter dominated universe~$[\Omega_{m}=1$, $\Omega_\phi=0$, $w_{\rm eff}=w$, $w_{\phi}=\frac{1}{2\alpha-1}]$ with eigenvalues of the Jacobian matrix  $\Big\{\lambda_1=\frac{3(1+w)}{2}, \lambda_2 = -3(\alpha-1) (1+w), \lambda_3=\frac{3(2\alpha w+2\alpha-w-3)}{2(2\alpha-1)}\Big\}$ of the perturbed matrix at $A_1$. Therefore, this point behaves as unstable node for $\frac{1}{2}+\frac{1}{w+1}<\alpha<1$, otherwise it behaves as a saddle. Thus, even though point $A_1$ lies on the intersection of all invariant sub-manifolds, it cannot be a global attractor.
        
        \item The set $A_2$ corresponds to an accelerated scalar field dominated solution~$[\Omega_{m}=0$, $\Omega_\phi=1$, $w_{\rm eff}=-1, w_\phi=-1]$. It is a non-hyperbolic set with eigenvalues $\Big\{\lambda_1=0, \lambda_2=-\frac{3}{2\alpha-1}, \lambda_3=-3(1+w)\Big\}$. However, as the dimension of the set is same as the number of vanishing eigenvalues, it is therefore a normally hyperbolic set \cite{coley}.  Therefore, points on this set behave as  stable points for $\alpha>\frac{1}{2}$ but this set cannot be global attractor as points on it corresponds to $y\neq 0$. 
    \end{itemize}
  Thus, in this case, the overall dynamics is very simple where the universe after evolving from matter domination (point $A_1$), it evolves towards an accelerated DE dominated solution  (set $A_2$). Hence, at the background level, this model can successfully explain the structure formation as well as the late time evolution of the universe for $\alpha>\frac{1}{2}$. This is more visible from  Fig. \ref{fig:Cos_Par1}, where we plot the time evolution of different energy density parameters along with the effective EoS against the redshift $z = \frac{a_0}{a}-1$ (with $a_0= 1$ the present scale factor) with $\alpha=2$. Here, while $z = 0$ (which corresponds to $a=a_0$)  represents the present time of the universe,  $z \approx  -1$ (which corresponds to $a \rightarrow \infty$) represents an asymptotic future of the universe.

    \begin{figure}
        \centering
        \subfigure[]{%
            \includegraphics[width=5cm,height=3cm]{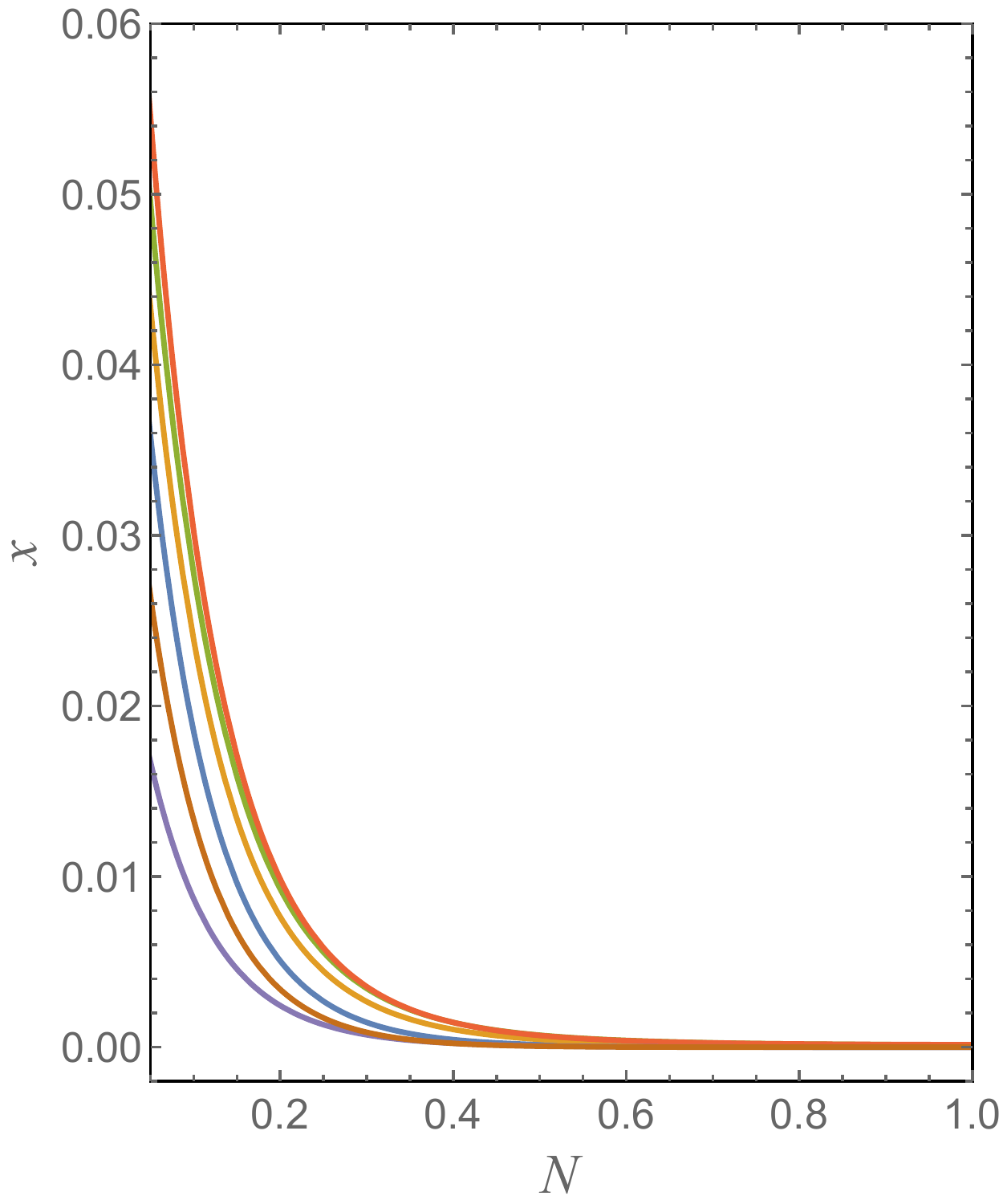}\label{fig:pert_c2_x}}
        \qquad
        \subfigure[]{%
            \includegraphics[width=5cm,height=3cm]{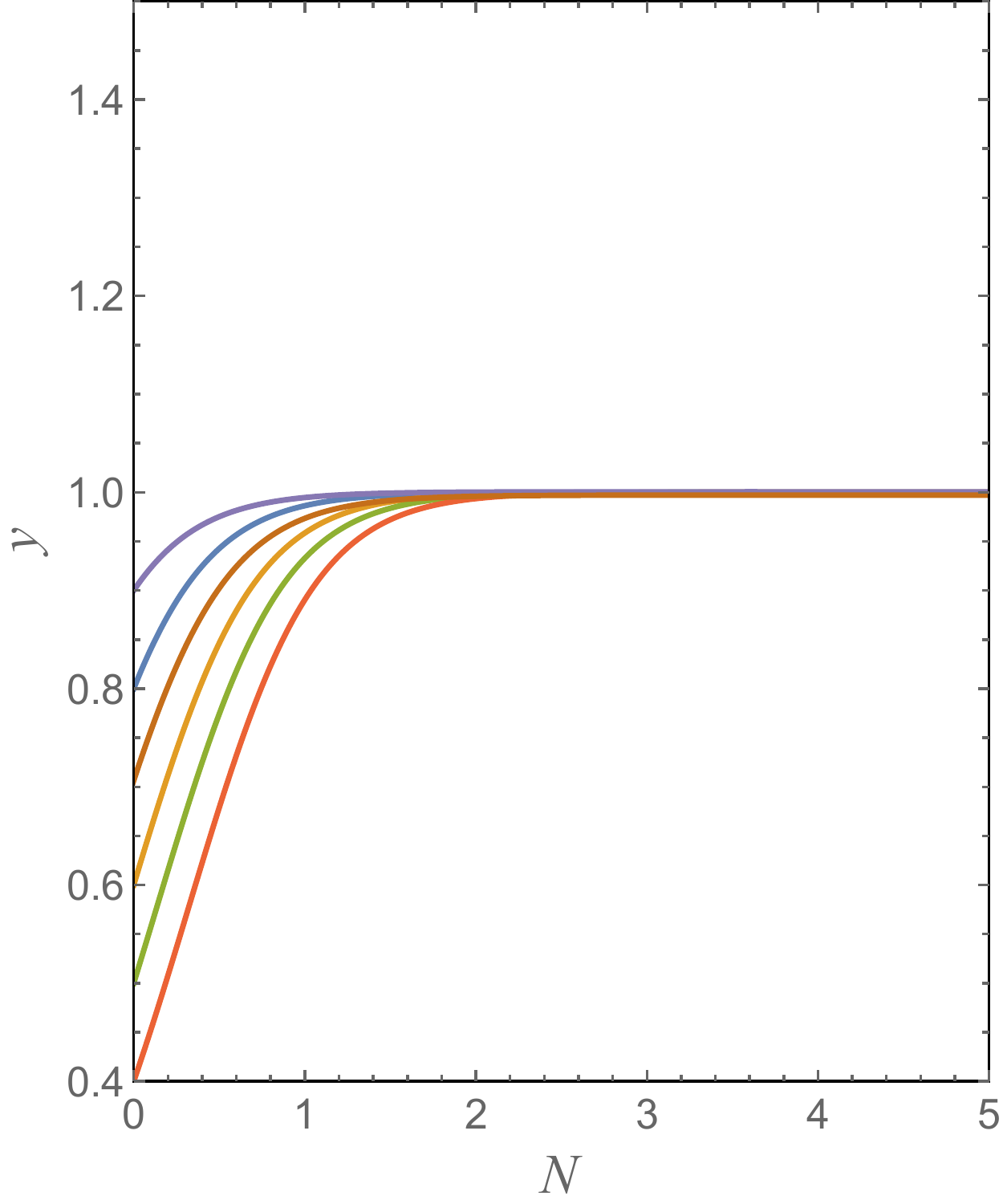}\label{fig:pert_c2_y}}
        \qquad
        \subfigure[]{%
            \includegraphics[width=5cm,height=3cm]{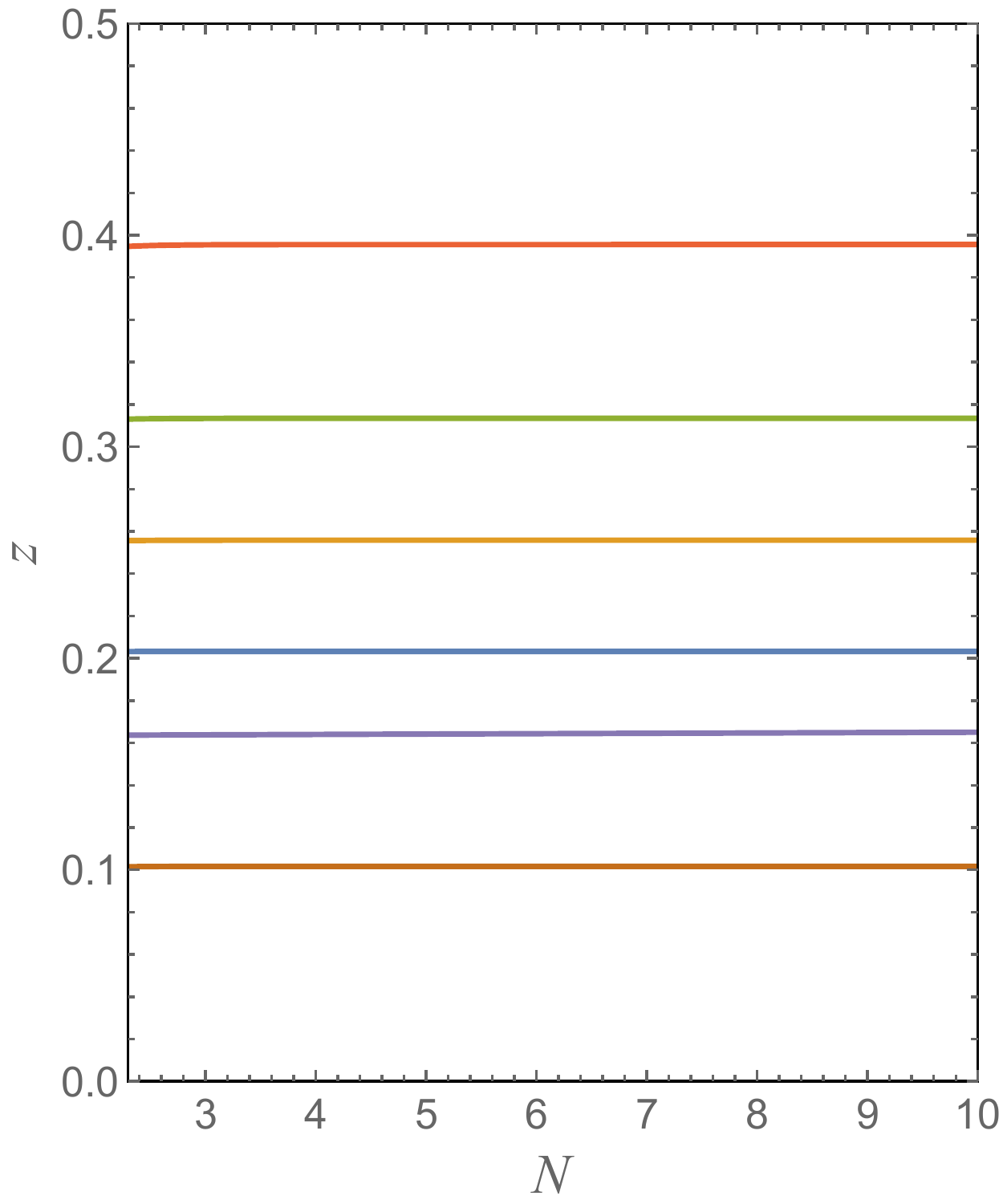}\label{fig:pert_c2_z}}
        \qquad
        \subfigure[]{%
            \includegraphics[width=5cm,height=3cm]{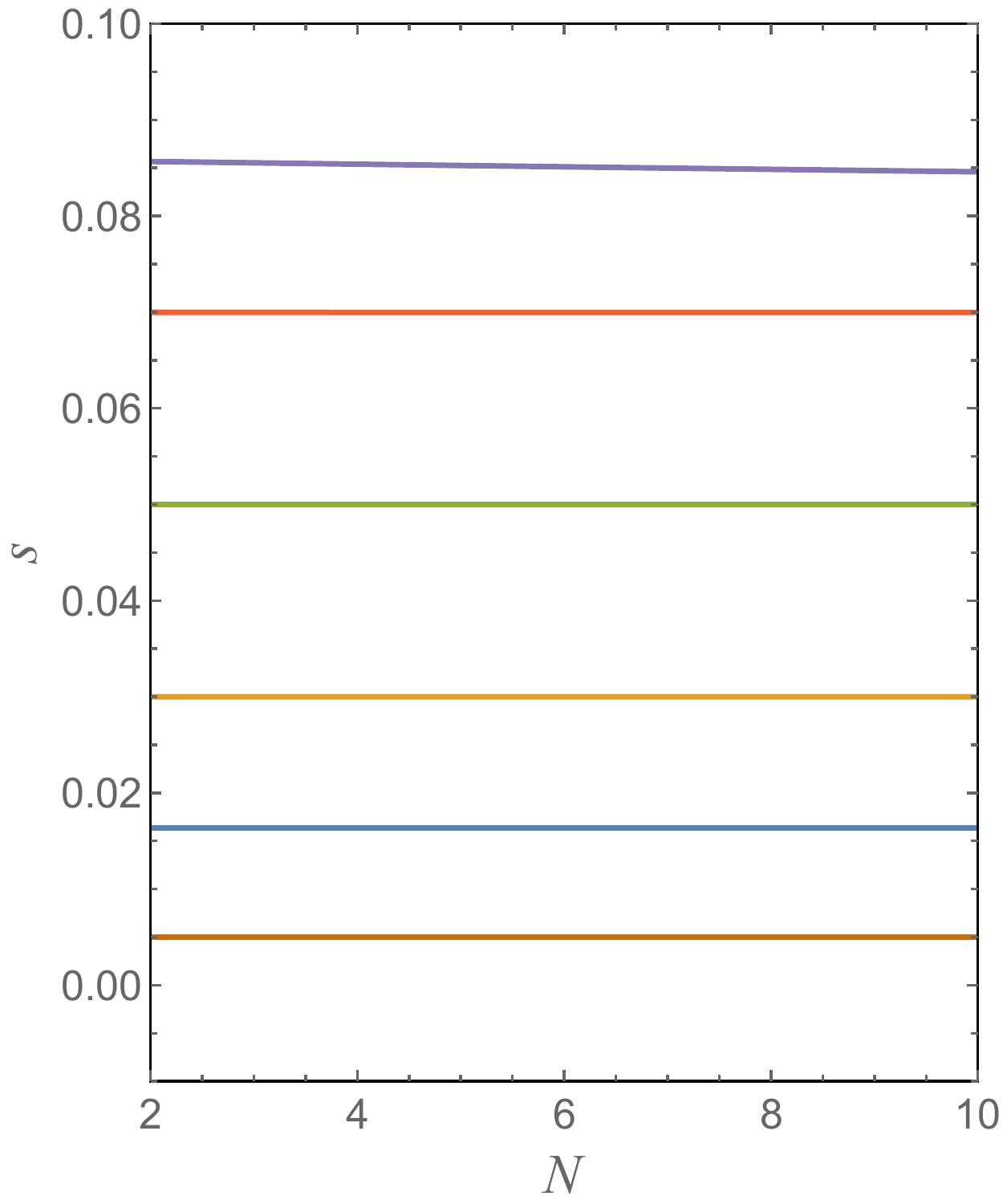}\label{fig:pert_c2_s}}
        \caption{Projection of the time evolution of phase space trajectories along the (a) $x$-axis, (b) $y$-axis, (c) $z$-axis and (d) $s$-axis which determine the stability of set $B_2$. Here we take $w=0$ and $\alpha=0.6$, $V=\frac{V_0}{\phi^n}$ with $n=4$ and $f=$ constant.}
        \label{fig:pert_c2}
    \end{figure} 
    
    \subsection{$ f(\phi)=$Constant}\label{f_constant}
    In this subsection, we investigate the case where only the function $f$ is constant. In this case variable $v$ vanishes and thus the system \eqref{xprime}-\eqref{vprime} reduces to a four-dimensional system in $x,y,z,s$. The system contains one invariant sub-manifold $y=0$ and one singular invariant sub-manifold $z=0$. Depending on the choice of potential, the system may possess another invariant sub-manifold $s=0$. For $\alpha=1$, we have checked that the behaviour of the system coincides with that of the canonical scalar field \cite{Copeland:1997et}, therefore, we will not further analyse this case. However for $\alpha \neq 1$, there are only two set of critical points: $B_1=(0, 0, z, s)$ and $B_2=(0, 1, z, s)$. Both sets are independent of the choice of the potential for their existence. From the second term of Eqs. (\ref{xprime}) and \eqref{cs2}, it can be seen that both sets demand  $ \alpha<1$ for their existence.

    \begin{itemize}
        \item The set $B_1$ corresponds to an unaccelerated DM dominated solution $[\Omega_{m}=1$, $\Omega_\phi=0$, $w_{\rm eff}=w, w_\phi=\frac{1}{2\alpha-1}]$. It is a normally hyperbolic set of points with eigenvalues $\Big\{\lambda_1=\frac{3(1+w)}{2}, \lambda_2= 3(1-\alpha)(1+w), \lambda_3=0, \lambda_4=\frac{3(2\alpha w+2\alpha-w-3)}{2(2\alpha-1)}\Big\}$. Therefore, points on this set are behaving as unstable node for $\frac{1}{2}+\frac{1}{w+1}<\alpha<1$, otherwise they are behaving as a saddle. 
        
        \item  The set $B_2$ corresponds to an accelerated scalar field dominated solution $[\Omega_{m}=0$, $\Omega_\phi=1$, $w_{\rm eff}=-1, w_\phi=-1]$. It is also non-hyperbolic in nature with  eigenvalues $\{\lambda_1=0, \lambda_2=0, \lambda_3=0, \lambda_4=-3(1+w)\}$ but not normally hyperbolic. Usually, the centre manifold theory is employed to determine the stability of points on this set \cite{wiggins}. However, in this work, we will simply refer to a numerical computational method of perturbation plot to determine the stability near points of this set. This numerical method has been found to be quite successful in literature for determining the stability behaviour of critical points of the system \cite{Dutta:2017fjw,Dutta:2017kch,Dutta:2017wfd,Dutta:2016bbs,Zonunmawia:2017ofc}. For this, we perturb the system near points of this set and plot the projections of phase trajectories along the $x$, $y$, $z$ and $s$ axes as given in Fig. \ref{fig:pert_c2}.  We note here that for these plots, we choose the values of initial conditions that are very close to the points of the set. It is evident from Figs. \ref{fig:pert_c2_x} and \ref{fig:pert_c2_y}, that the phase space trajectories approach  $x=0$ and $y=1$ respectively as $N \rightarrow \infty$. Further, it can be seen in Fig. \ref{fig:pert_c2_z}, \ref{fig:pert_c2_s}, that trajectories starting from any values of $z$ and $s$ remains almost constant. From the behaviour of the perturbed system near the set $B_2$, we can conclude that the set $B_2$ behaves as a stable set but points on this set cannot be global attractors as they do not lie on the intersection of invariant sub-manifolds. 
    \end{itemize}
    Hence, for a restricted choice of $\alpha$ ($\frac{1}{2}<\alpha<1$), this class of models describes the simple evolution of the universe where after the universe passes through matter domination (set $B_1$), it eventually evolves towards an accelerated DE dominated solution  (set $B_2$).

    \subsection{$V(\phi)=$Constant} \label{v_constant}
\begin{figure}
        \centering
        \subfigure[]{%
            \includegraphics[width=6cm,height=5cm]{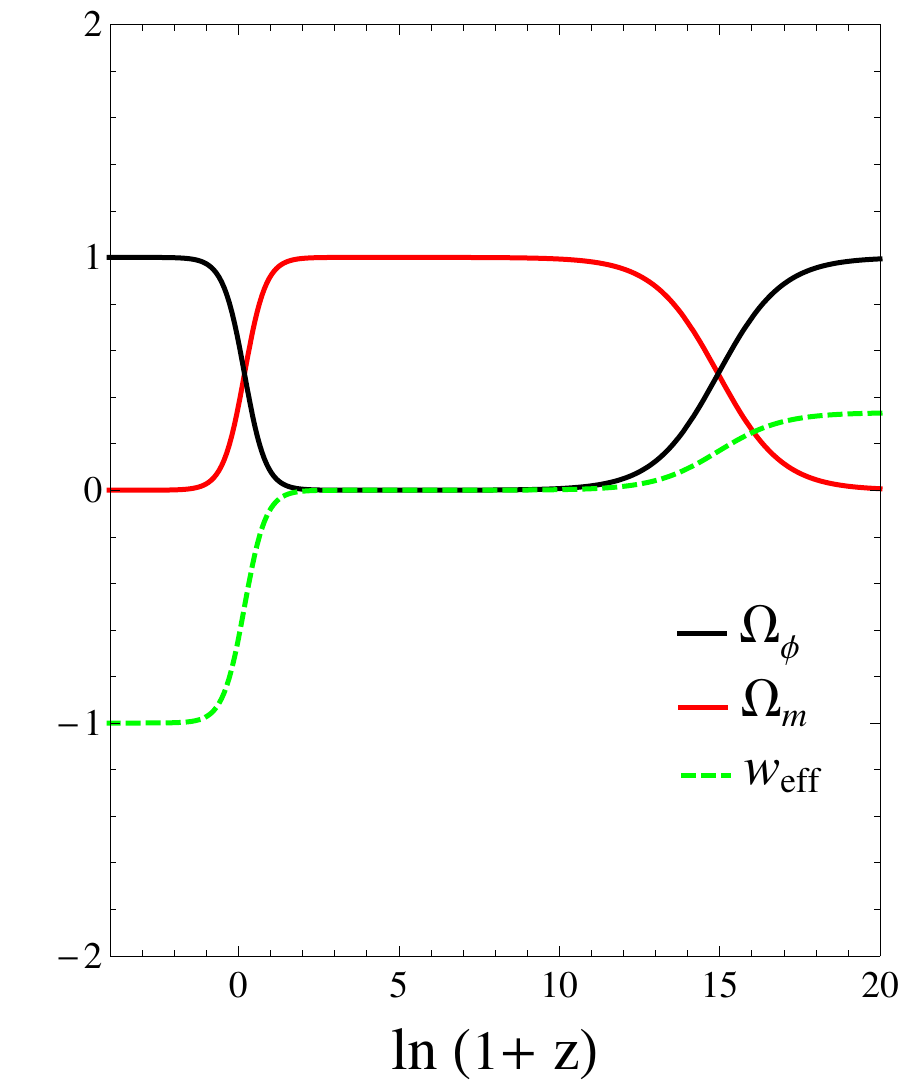}\label{Cos_Par_2}}
        \qquad
        \subfigure[]{%
            \includegraphics[width=6cm,height=5cm]{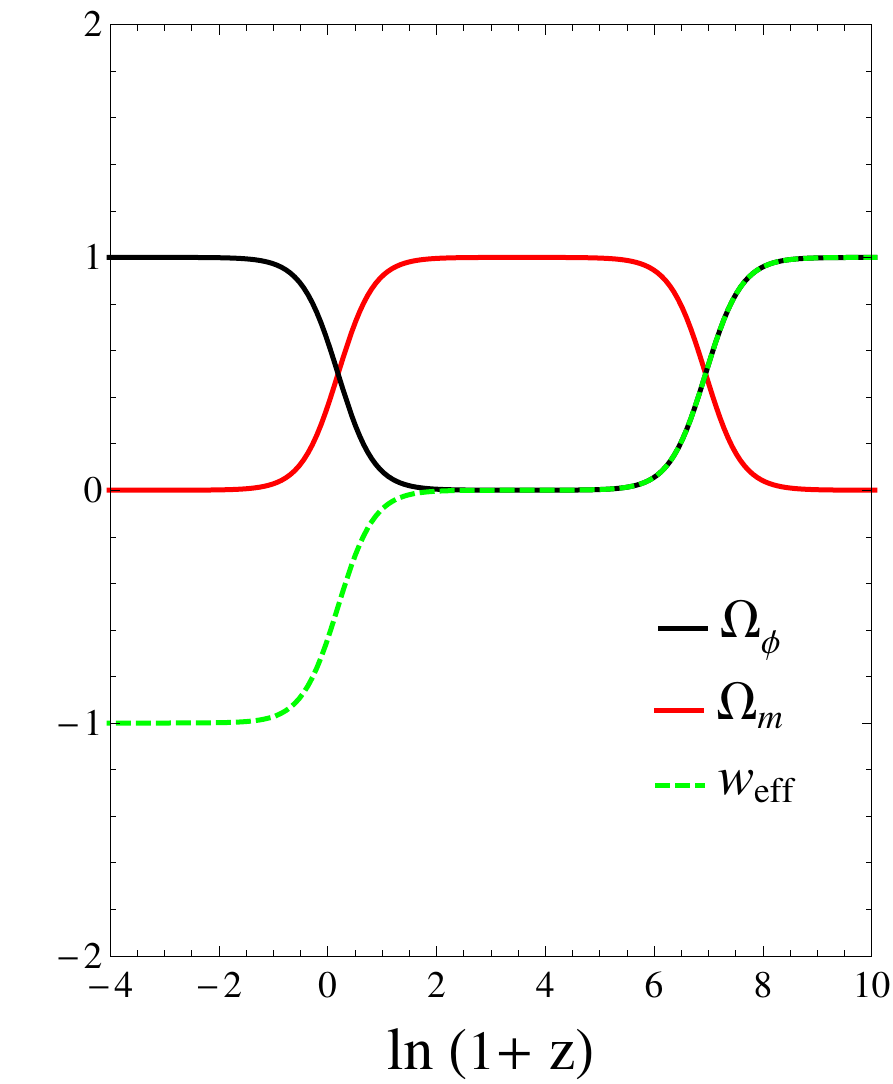}\label{fig:Cos_Parr2}}
        \caption{The time evolution of the relative matter energy density $\Omega_m$, the relative scalar field energy density $\Omega_\phi$ and the effective EoS $w_{\rm eff}$. Here we take $f(\phi)=f_0 e^{\beta \phi}$, $\beta=5$ and $V=$constant, $w=0$ with $\alpha=2$ in (a) and $\alpha=1$ in (b).}\label{fig:Cos_Par2}
    \end{figure}

    In this subsection, we investigate the case where only the potential is constant. For this choice of potential, the variable $s$ vanishes, so the system \eqref{xprime}-\eqref{vprime} reduces to a four dimensional system in $x, y, z$ and $v$. The system thus contains three invariant sub-manifolds $x=0, y=0$ and $z=0$, however, depending on the choice of $h(v)$, the system may contains another invariant sub-manifold $v=0$. Under this case, we have two sets of critical points $C_1=(0, 0, 0, v)$,  $C_2=(0, 1, z, v)$ and three critical points  $C_3=\Big(-\frac{\alpha\sqrt{6}(2\alpha w+2\alpha-w-3)}{2v_*(2\alpha-1)}, 0, 0, v_*\Big)$, \\ $C_4=\Big(\frac{\sqrt{6}\alpha}{v_*(2\alpha-1)}, 1, 0, v_*\Big)$ and  $C_5=\Bigg(-\frac{\alpha \sqrt{6}(\alpha-1)}{v_*(2\alpha-1)}, 0, $\\ $\frac{3^{(1-\alpha)}\Big[-\frac{\alpha \sqrt{6}(\alpha-1)}{v_*(2\alpha-1)}\Big]^{-2\alpha}}{(2\alpha-1)}, v_*\Bigg)$. Note that from the second term of the equation \eqref{xprime}, the existence of point $C_4$ specifically demands $s=0$ i.e. the constant potential only. Further, points $C_3$, $C_4$, $C_5$ do not exists when $v_*=0$. 
    \begin{itemize}
        \item The set of critical points $C_1$ corresponds to a matter dominated universe~$[\Omega_{m}=1$, $\Omega_\phi=0$, $w_{\rm eff}=w, w_\phi=\frac{1}{2\alpha-1}]$ with eigenvalues  $\Big\{\lambda_1=\frac{3(1+w)}{2}, \lambda_2=-3(\alpha-1)(1+w), \lambda_3=\frac{3(2\alpha w+2\alpha-w-3)}{2(2\alpha-1)}, \lambda_4=0 \Big\}$. Therefore, points on this set behaving as unstable node for  $\frac{1}{2}+\frac{1}{w+1}<\alpha<1$, otherwise they are behaving as  saddle points. This again implies that points on this set cannot be global attractor.
        
        \item The set $C_2$ corresponds to an accelerated scalar field potential dominated solution $[\Omega_{m}=0$, $\Omega_\phi=1$, $w_{\rm eff}=-1, w_\phi=-1]$. It is a normally hyperbolic set with eigenvalues $\Big\{\lambda_1=0, \lambda_2=0, \lambda_3=-3(1+w), \lambda_4=-\frac{3}{2\alpha-1}\Big\}$. Therefore, it is always stable but it cannot be global attractor as $y\ne 0$.
        
        \item The point $C_3$ corresponds to a matter dominated solution with $[\Omega_{m}=1$, $\Omega_\phi=0$, $w_{\rm eff}=w, w_\phi=\frac{1}{2\alpha-1}]$. The eigenvalues of the corresponding perturbed matrix are \\ $\Big\{\lambda_1=\frac{3(1+w)}{2},~~ \lambda_2=\frac{3(2\alpha w -w-1)}{2\alpha-1}, ~\lambda_3=-\frac{3(2\alpha w+2\alpha-w-3)}{2(2\alpha-1)},~\\ \lambda_4= \frac{3\alpha(2\alpha w +2\alpha-w-3)dh(v_*)}{v_*(2\alpha-1)} \Big\}$. Since $\lambda_1>0$ but $\lambda_2$, $\lambda_3$ are not both positive, therefore this point is a saddle for any choice of function $f$.
        
        \item The point $C_4$ corresponds to an accelerated  solution dominated by the potential term of the scalar field $[\Omega_m=0, \Omega_\phi=1, w_{\rm eff}=-1, w_\phi=-1 ]$ with eigenvalues $\Big\{\lambda_1=-3(1+w), \lambda_2= \frac{3}{2\alpha-1}, \lambda_3=-\frac{6\alpha}{2\alpha-1}, \lambda_4= -\frac{6\alpha dh(v_*)}{v_*(2\alpha-1)}\Big\}$. It is therefore saddle for $\alpha>0$ and also for $\alpha<0$ with $v_* dh(v_*)<0$, however, it is stable for $\alpha<0$ and $v_* dh(v_*)>0$. This point corresponds to a slow roll behaviour where potential part of the scalar field dominates and its derivative vanishes (as its existence demands $s=0$).  Its saddle property is interesting as it provides an explanation of how the Universe could possibly evolve towards a slow roll regime  and eventually escape. It is also worth to note that this type of critical point is also obtained in \cite{DeSantiago:2012nk}.
        
        \item The critical point $C_5$ corresponds to a solution dominated by the kinetic term of a scalar field   $[ \Omega_m=0, \Omega_\phi=1,  w_{\rm eff}=\frac{1}{2\alpha-1}, w_{\phi}=\frac{1}{2\alpha-1} ]$. The eigenvalues of the corresponding perturbed matrix are $\Big\{\lambda_1=\frac{3 \alpha}{2\alpha-1}, \lambda_2= \frac{3 (1-\alpha)}{2\alpha-1},$\\$ \lambda_3=\frac{3\left(1-w(2\alpha-1)\right)}{2\alpha-1}, \lambda_4= \frac{6\alpha (\alpha-1)}{2\alpha-1} \frac{dh(v_*)}{v_*}\Big\}$. This point is an unstable node if  $\frac{1}{2}<\alpha<1$ and $v_*\,dh(v_*)<0$, otherwise it is a saddle. It is worth mentioning that the existence of this point is cosmologically very interesting as it describes the  early radiation domination for $\alpha=2$, the stiff matter solution for $\alpha=1$ and  the matter domination for very large values of $\alpha$.
    \end{itemize}
  Hence, from the above analysis it can be seen that for those  function $f$ in which $v_*\neq 0$, the system evolves from the point $C_5$ which behaves as a radiation (i.e. $w_{\rm eff}=\frac{1}{3}$ for $\alpha=2$) or DM  ($w_{\rm eff}\approx 0$ for large $\alpha$) or stiff matter ($w_{\rm eff}=1$ for $\alpha=1$) towards matter domination describe by either a set $C_1$ or point $C_3$ with $w_{\rm eff}=0$ and then eventually settle in a DE dominated epoch described by a set $C_2$ with $w_{\rm eff}=-1$.  Therefore, a viable sequence of cosmological eras (radiation $\rightarrow$ matter $\rightarrow$ dark energy) is achieved by choosing the trajectory: $C_5\,\rightarrow C_1\,\rightarrow C_2$ or $C_5\,\rightarrow C_3\,\rightarrow C_2$ for $\alpha=2$.  Further, the early times radiation (for $\alpha=2$) or the DM behaviour (for large $\alpha$) is specifically due to the non-canonical scalar field. This explains the well-known unified  DE-DM behaviour in non-canonical setting where the kinetic term of the scalar field plays the role of DM and the potential plays the role of DE \cite{Kamenshchik:2001cp,Padmanabhan:2002cp,Scherrer:2004au,DiezTejedor:2006qh,DeSantiago:2011qb,Sahni:2015hbf,Mishra:2018tki}. Similar unified description is obtained in the context of generalized Galileon theories but without any instabilities at the perturbative level at all times \cite{Koutsoumbas:2017fxp}. In order to clearly see the background cosmological sequence in detail, we plot the time evolution of different relative energy densities along with the overall effective EoS by considering a specific choice of the function $f$ viz. $f=f_0 e^{\beta\phi}$ with $\beta=5$ and $\alpha=1, 2$ (see Fig. \ref{fig:Cos_Par2}). It is worth mentioning that this choice of $f$ is specifically taken so that the function $\Gamma_2$ can be written as function of $v$. Moreover, it has been found that this choice of $f$ fit with data from the combination of various datasets such as  SNIa, BAO and CMB \cite{Mamon:2015qai}. From Fig. \ref{fig:Cos_Par2}, it can be seen that the background evolution of the Universe resembles that of the $\Lambda$CDM model, with an early time contribution arising from the non-canonical scalar field term. However, the presence of a non-canonical parameter $\alpha$ in the speed of sound produces a possible deviation of this model from $\Lambda$CDM at the perturbation level \cite{Sahni:2015hbf}.

    \begin{figure}
        \centering
        \subfigure[]{%
            \includegraphics[width=5cm,height=3cm]{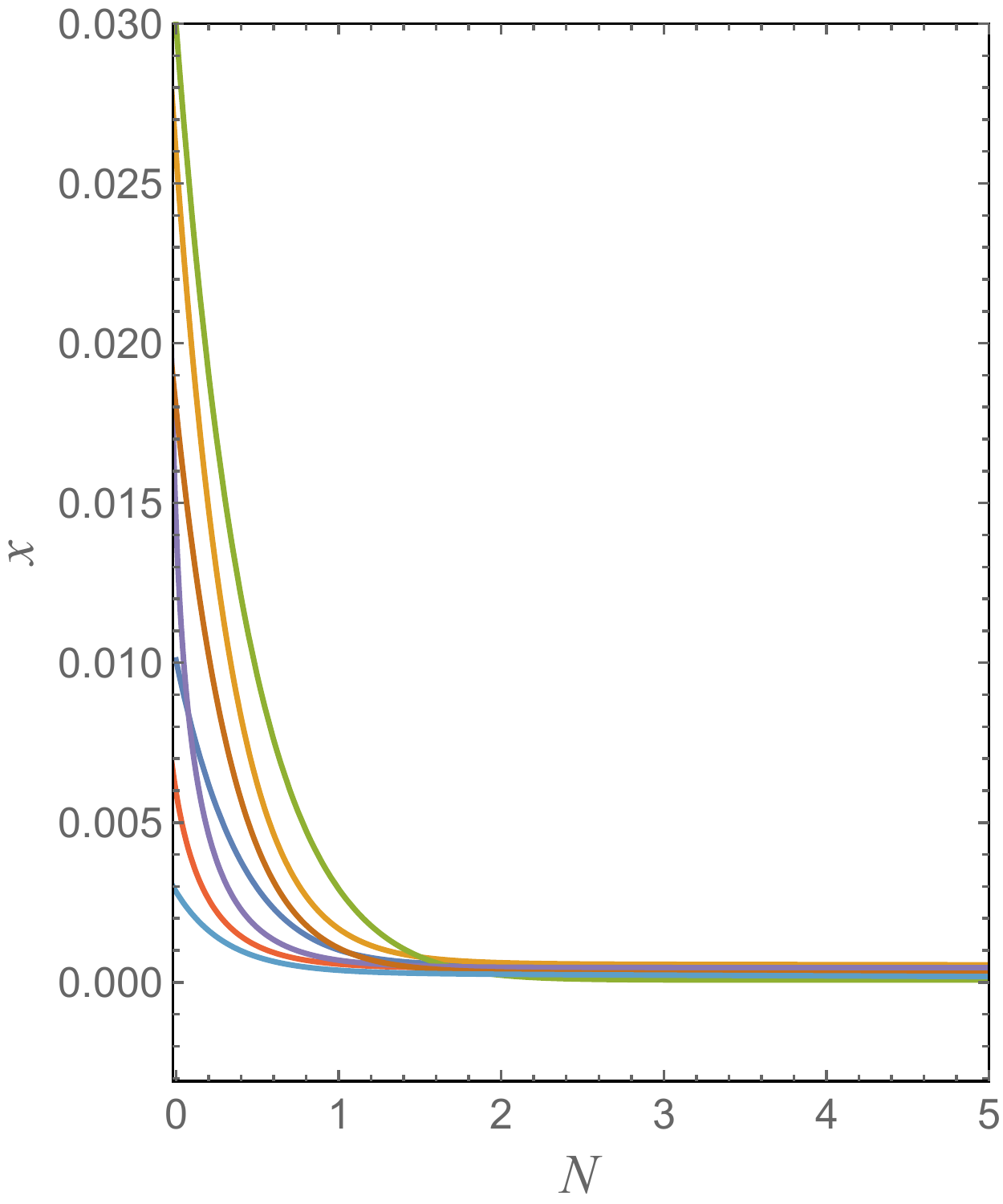}\label{fig:pertt_d2_x}}
        \qquad
        \subfigure[]{%
            \includegraphics[width=5cm,height=3cm]{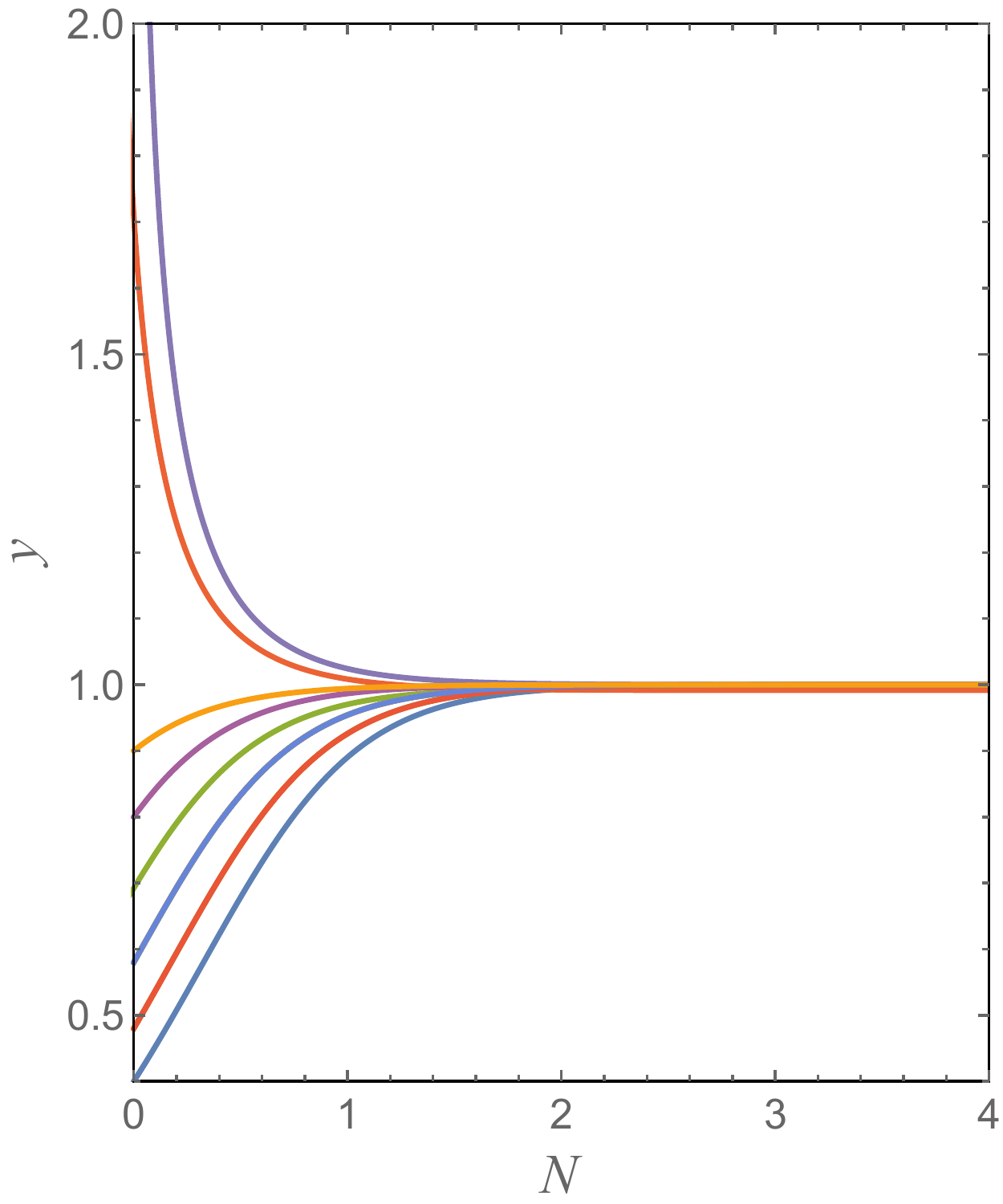}\label{fig:pertt_d2_y}}
        \qquad
        \subfigure[]{%
            \includegraphics[width=5cm,height=3cm]{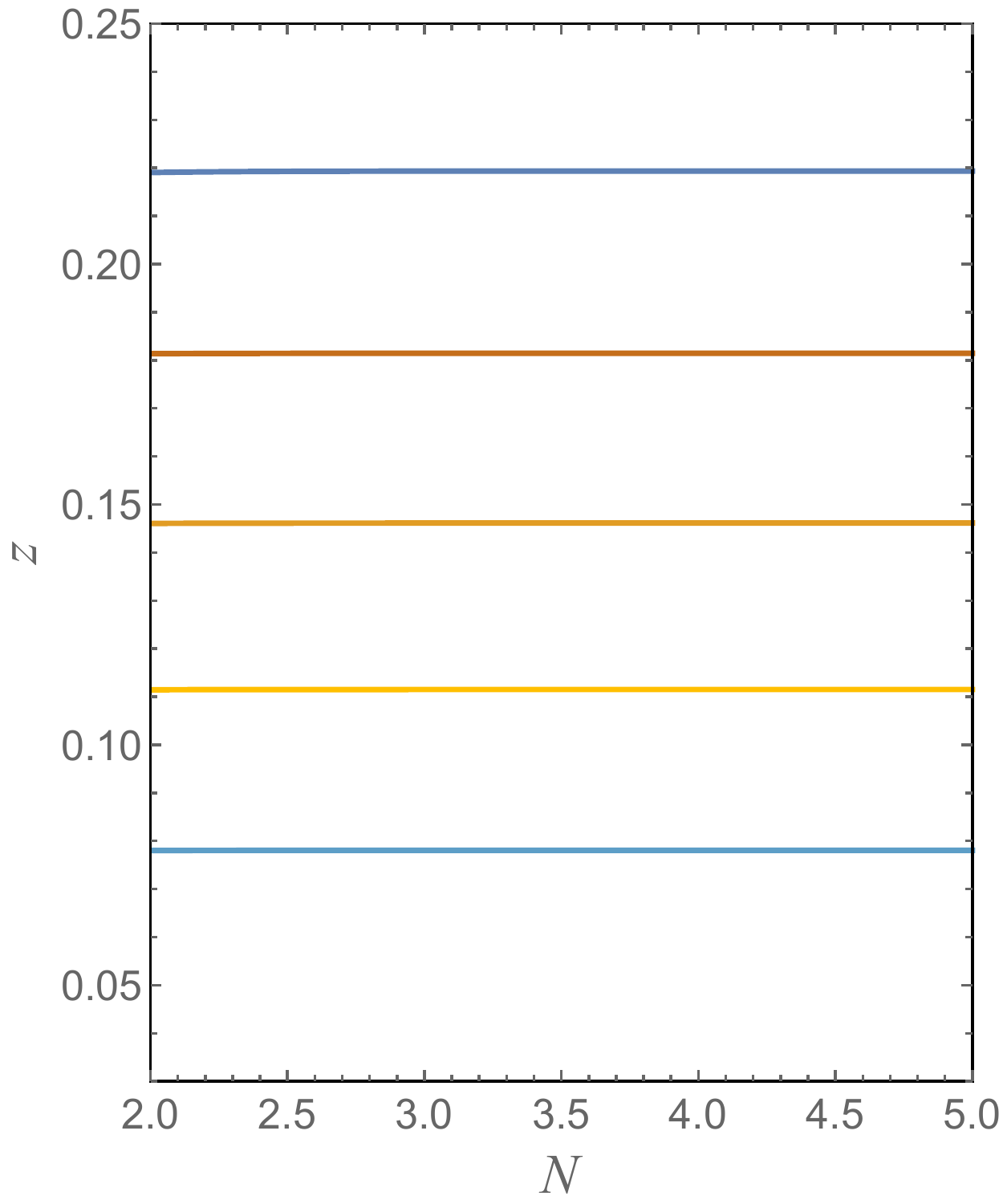}\label{fig:pertt_d2_z}}
        \qquad
        \subfigure[]{%
            \includegraphics[width=5cm,height=3cm]{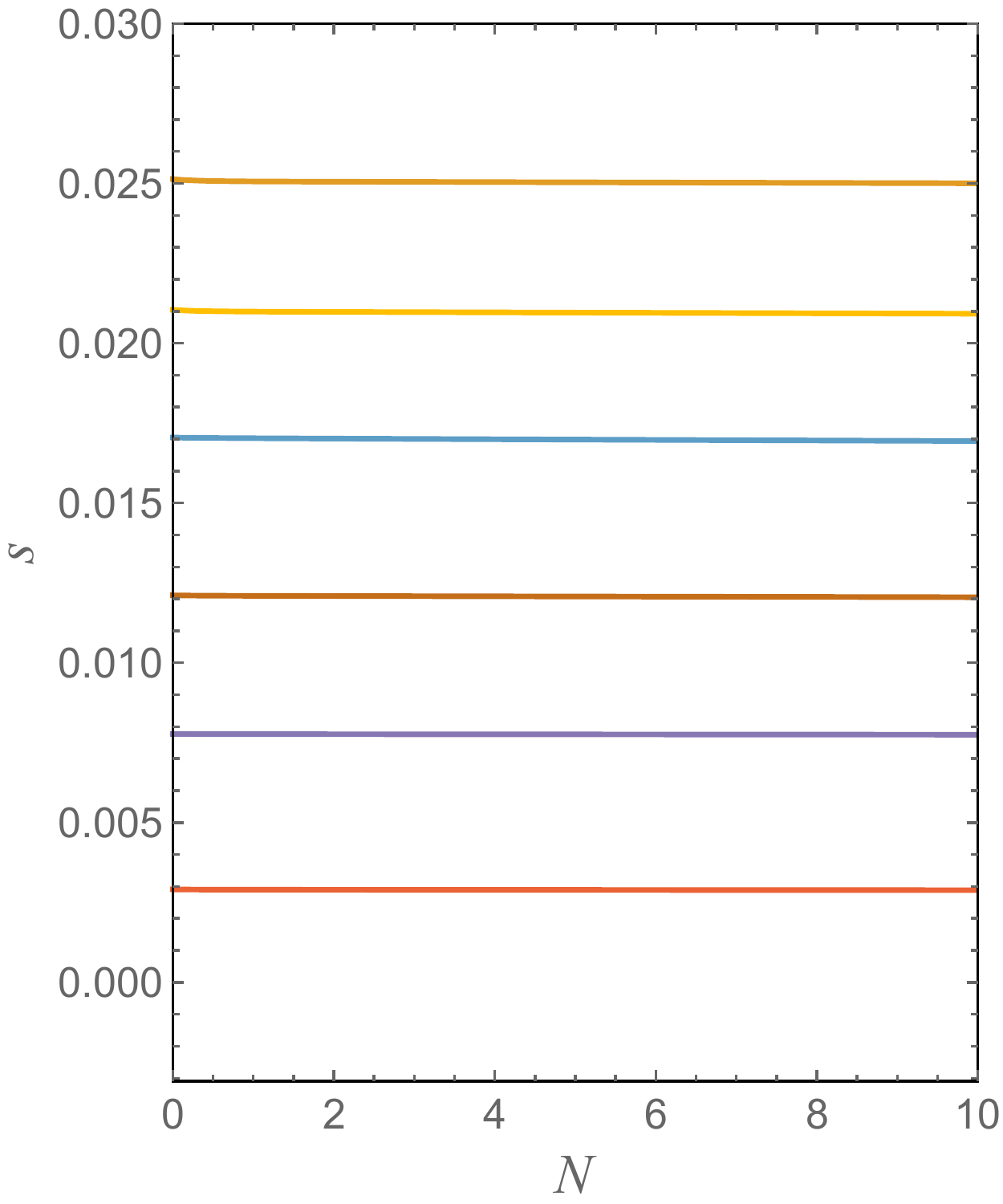}\label{fig:pertt_d2_s}}
        \qquad
        \subfigure[]{%
            \includegraphics[width=5cm,height=3cm]{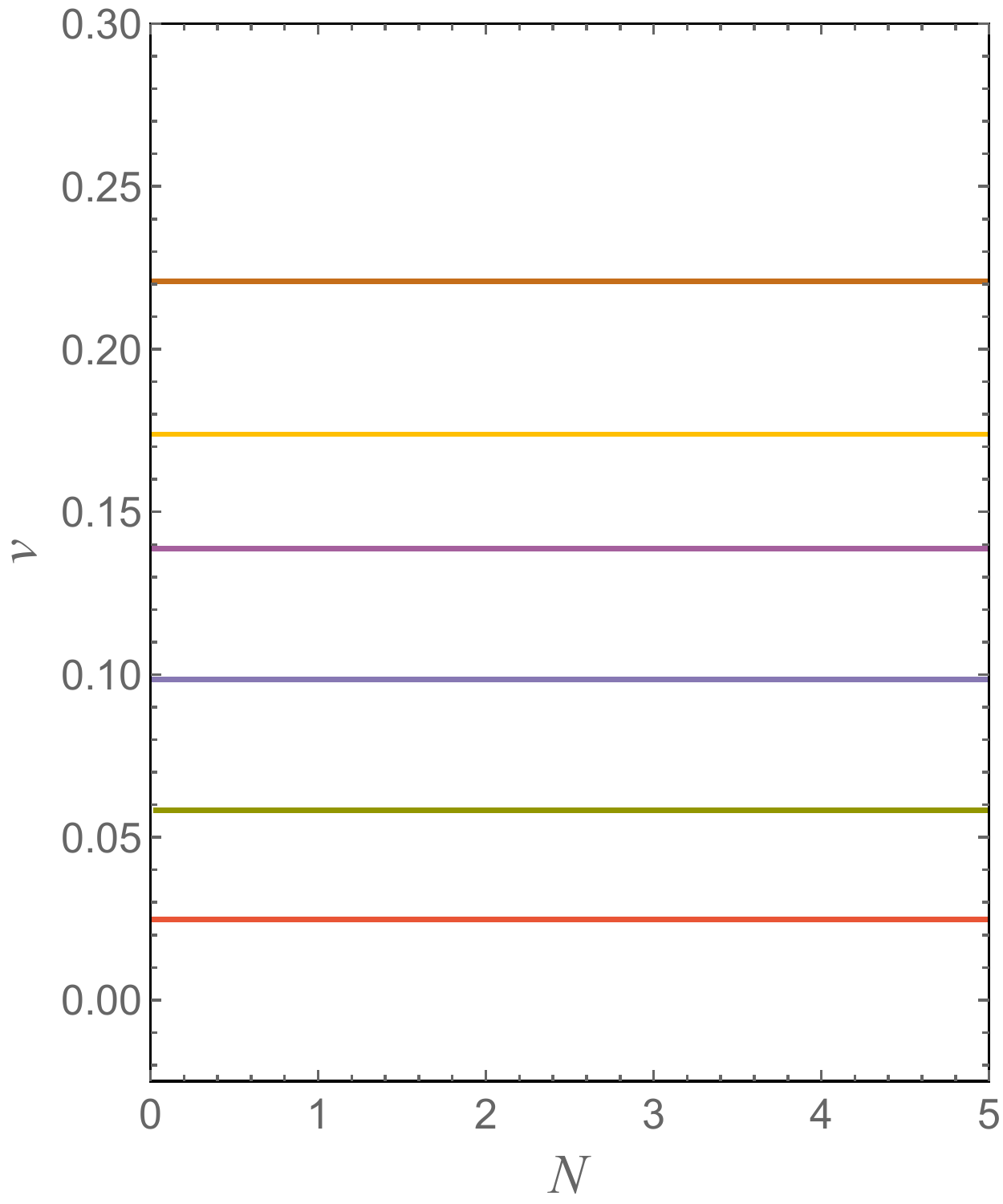}\label{fig:pertt_d2_v}}
        \caption{Projection of the time evolution of phase space trajectories along the (a) $x$-axis, (b) $y$-axis, (c) $z$-axis, (d) $s$-axis and (e) $v$-axis which determine the stability of set $D_2$. Here, we take $w=0$, $\alpha=0.8$, $V(\phi)=V_0(1-e^{-\lambda \phi})^2$ with $\lambda=4$ and  $f(\phi)=f_0 \phi^{-n}$ with $n=2$.}
        \label{fig:pertt_d2}
    \end{figure}

    \begin{figure}
        \centering
        \subfigure[]{%
            \includegraphics[width=6cm,height=5cm]{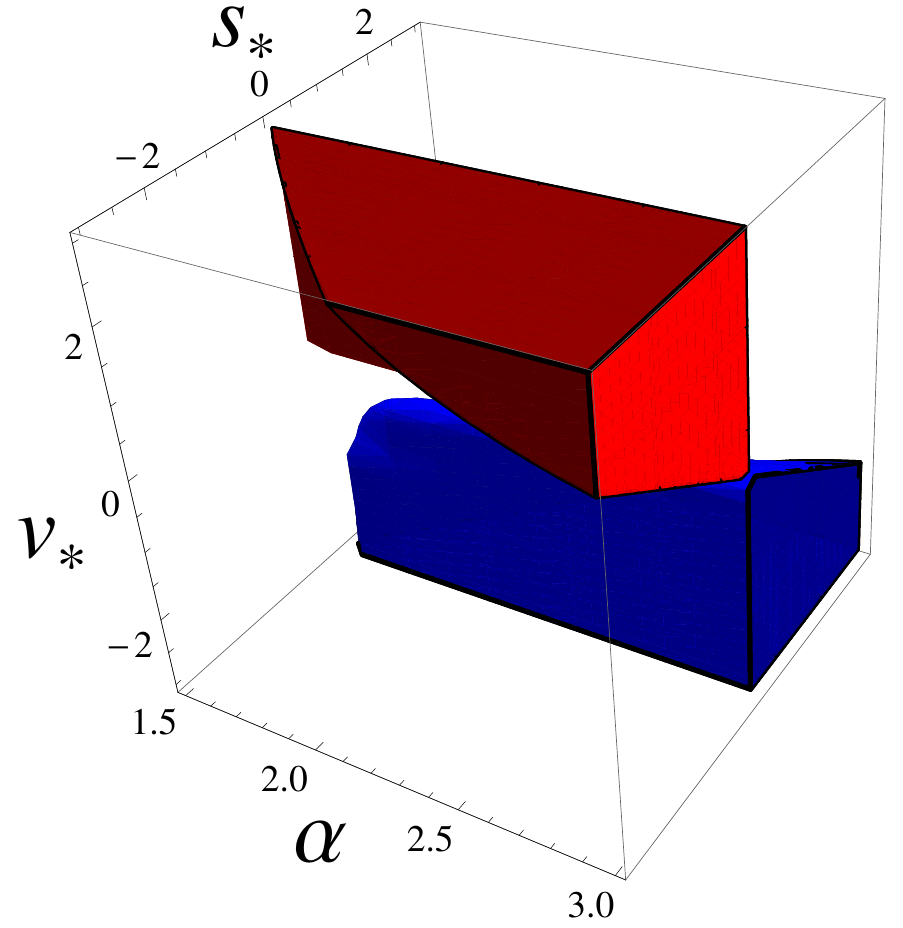}\label{fig:Region_Plot_D3_3D}}
        \qquad
        \subfigure[]{%
            \includegraphics[width=6cm,height=5cm]{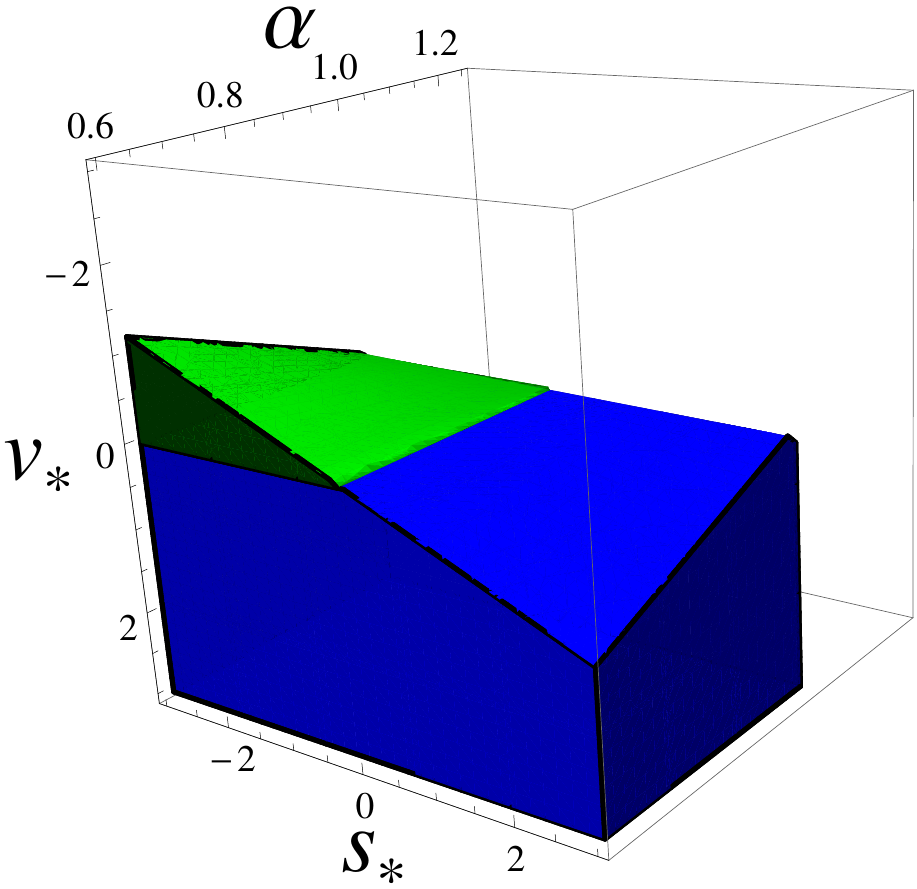}\label{fig:RegionPlotD4}}
        \caption{(a) {\it Regions of stability of point $D_3$}: The red shaded region represents the region of stability for $dg(s_*)>0$ and $dh(v_*)>0$, the blue shaded region of stability for the case when $dg(s_*)<0$ and $dh(v_*)<0$.  (b) {\it Regions of instability of point $D_4$}: The green shaded region represents the region of unstable node for $dg(s_*)>0$ and $dh(v_*)>0$, the blue shaded region represents the region of unstable node for the case when $dg(s_*)<0$ and $dh(v_*)<0$. In both panels, the remaining unshaded regions represent regions where points are saddle. Here, we have taken $w=0$.}  \label{fig:RegionPlot}
    \end{figure}
    
    \begin{figure}
        \centering
        \subfigure[]{%
            \includegraphics[width=5cm,height=3cm]{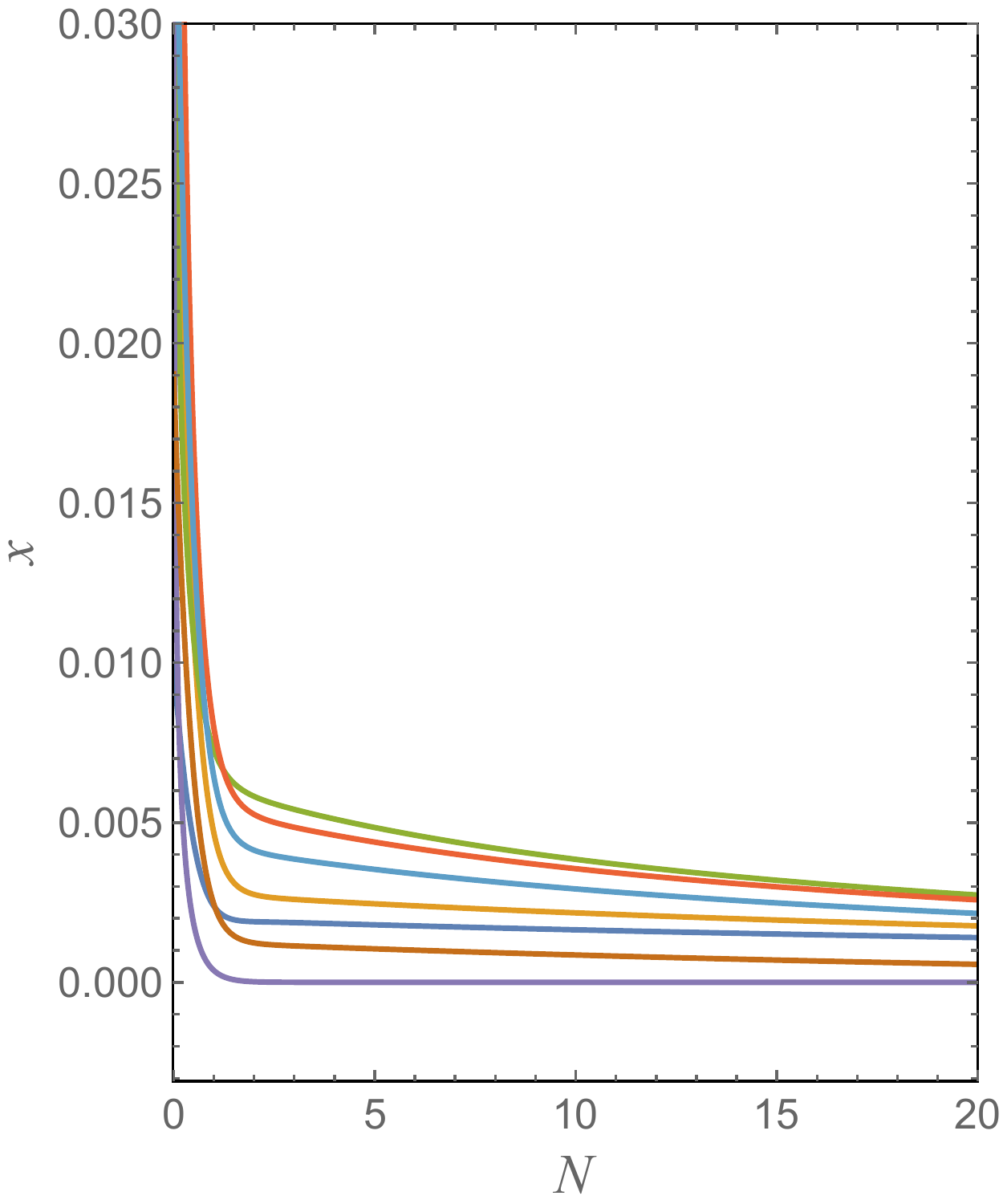}\label{fig:pert_e2_x}}
        \qquad
        \subfigure[]{%
            \includegraphics[width=5cm,height=3cm]{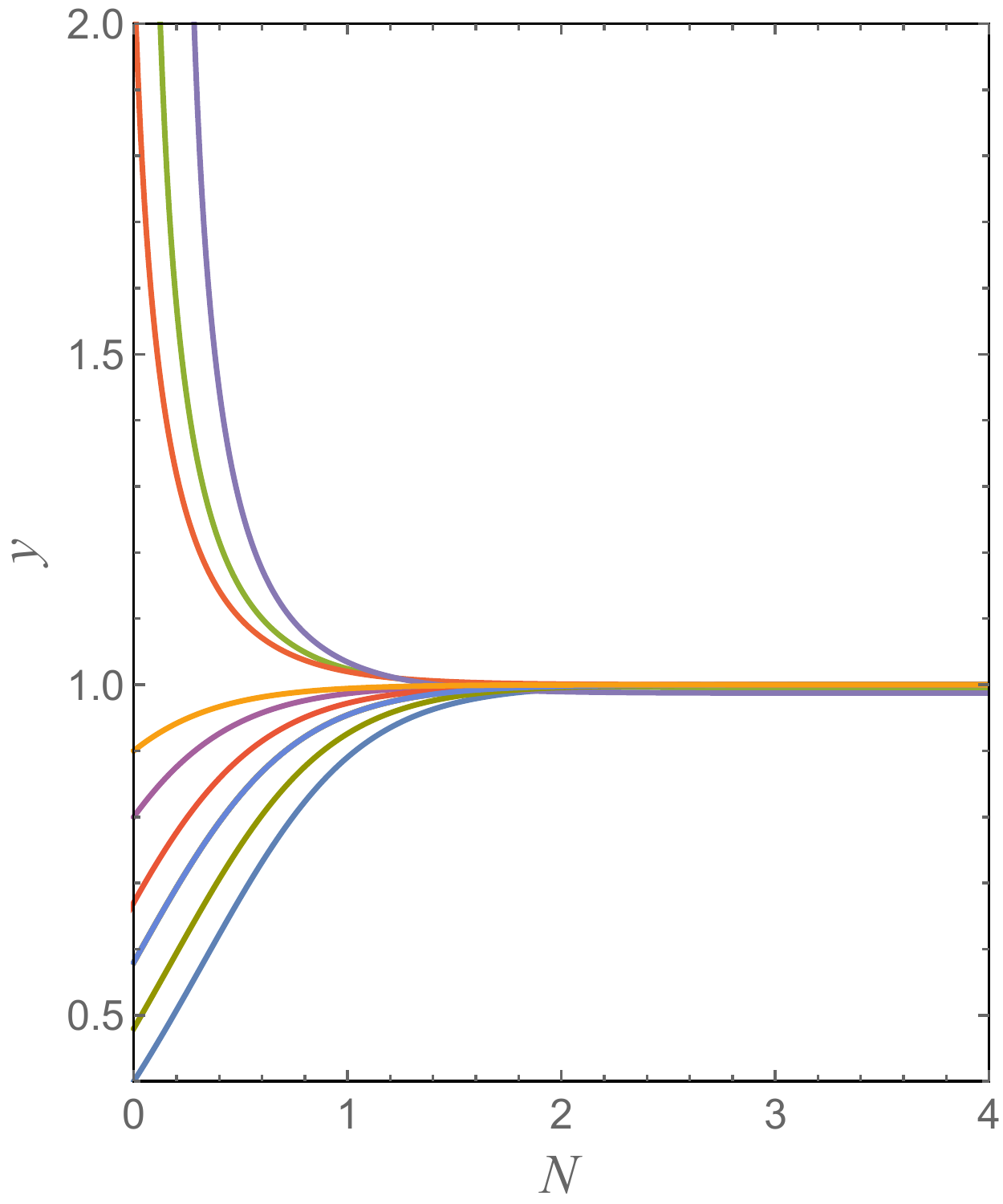}\label{fig:pert_e2_y}}
        \qquad
        \subfigure[]{%
            \includegraphics[width=5cm,height=3cm]{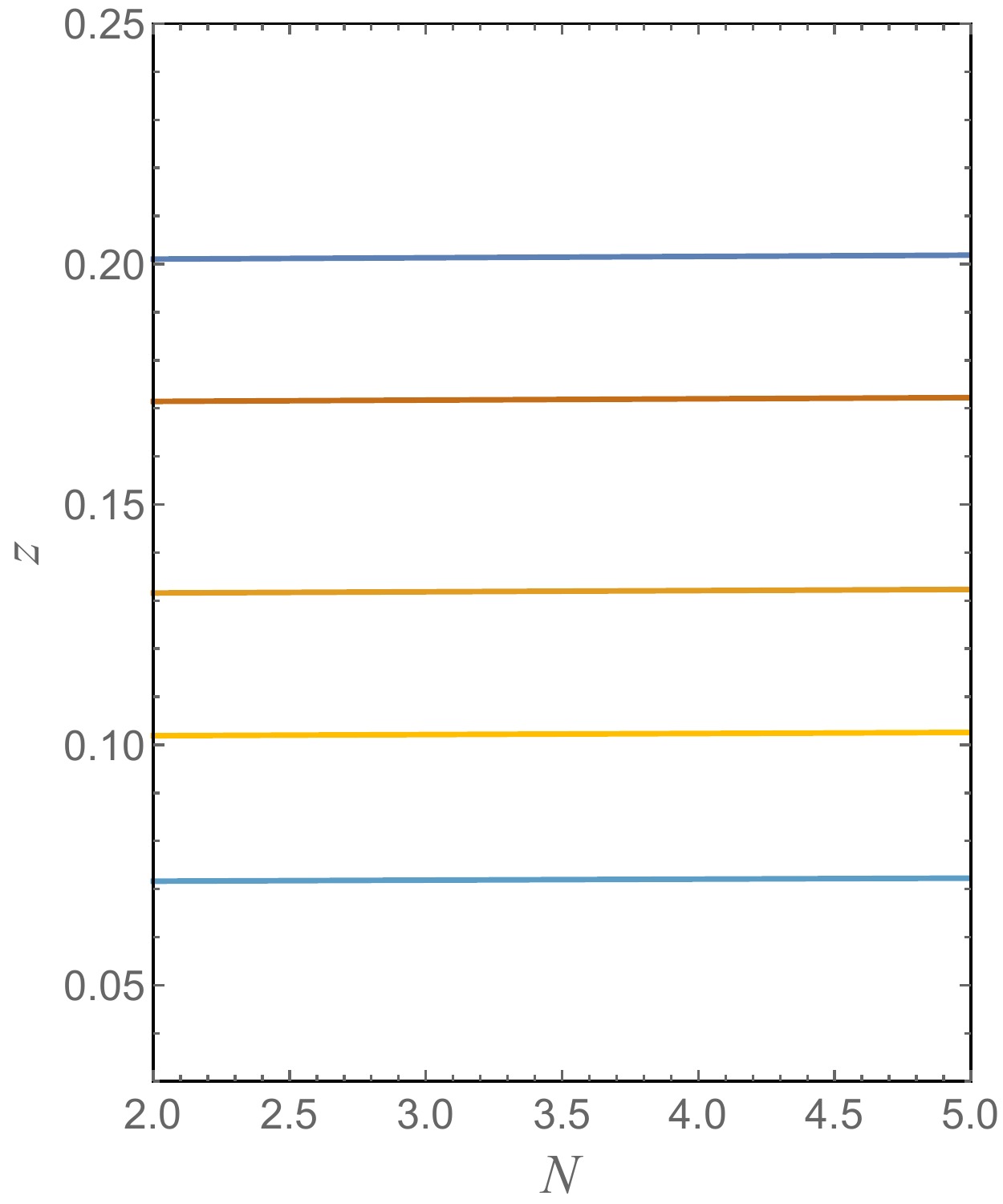}\label{fig:pert_e2_z}}
        \qquad
        \subfigure[]{%
            \includegraphics[width=5cm,height=3cm]{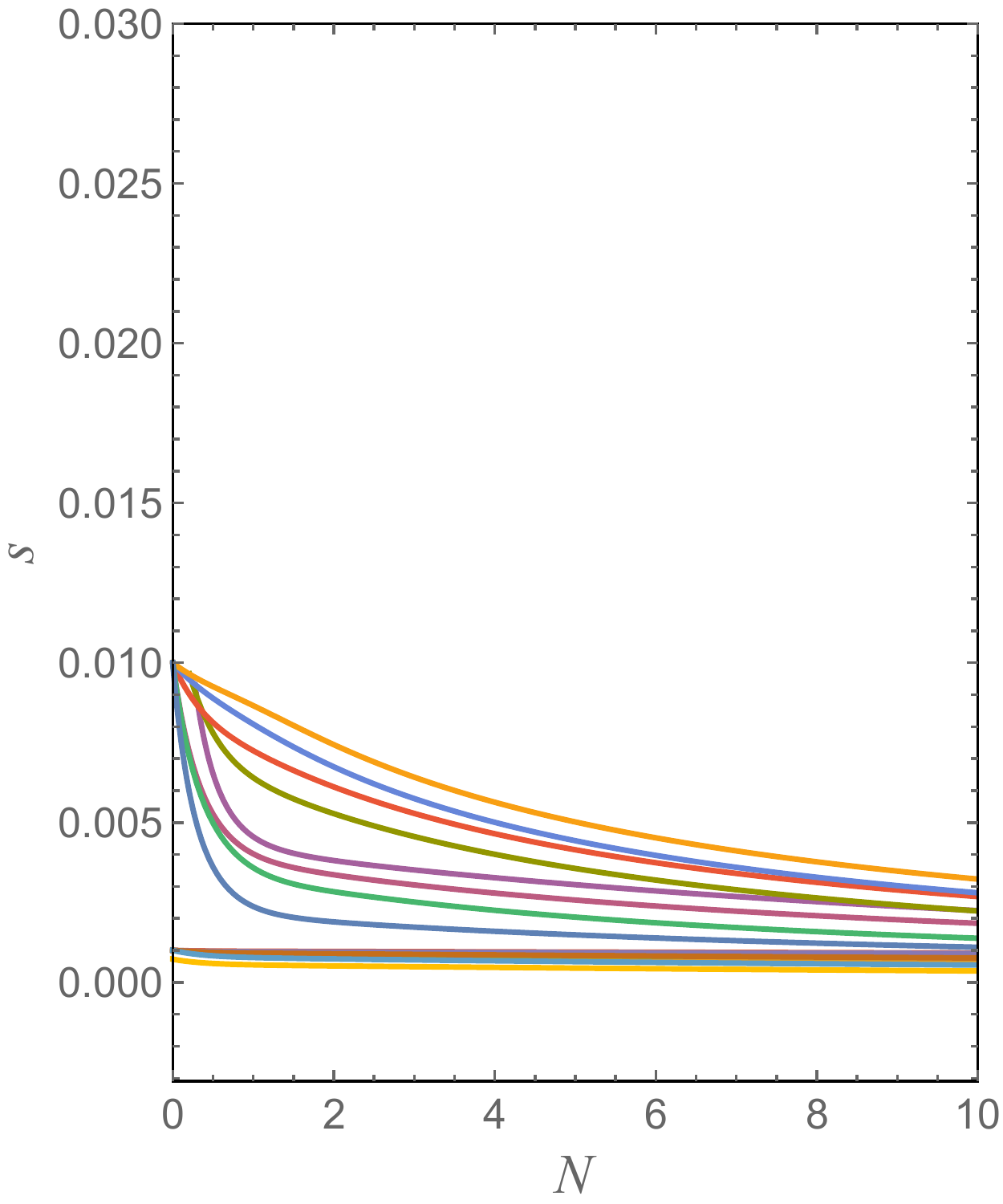}\label{fig:pert_e2_s}}
        \caption{Projection of the time evolution of phase space trajectories along the (a) $x$-axis, (b) $y$-axis, (c) $z$-axis and (d) $s$-axis which determine the stability of set $E_2$. Here we take $w=0$ and $\alpha=1$, $V=V_0 \phi^{-n}$ with $n=4$ and  $f(\phi)=f_0 e^{\beta \phi}$ with $\beta=2$.}
        \label{fig:pert_e2}
    \end{figure} 
    
    \subsection{$f(\phi)=$General function, $V(\phi)=$General potential}\label{f_gen_v_gen}
    In this subsection, we shall investigate the case where both function $f$ and potential $V$ are kept general. By general, we mean $V$ and $f$ are such that functions $\Gamma_1$, $\Gamma_2$ can be written as functions of $s$ and $v$ respectively. It is worth mentioning that constant potential and constant function $f$ do not belongs to this category as both $\Gamma_1$ and $\Gamma_2$ cannot be defined. For this case, the system possess an invariant sub-manifold $y=0$ and a singular invariant sub-manifold $z=0$. For some choice of $V$ and $f$, $s=0$ and $v=0$ may be invariant sub-manifolds. In what follows, we split the analysis of system \eqref{xprime}-\eqref{vprime} into two cases $\alpha\neq 1$ and $\alpha=1$.
    \subsubsection{$\alpha \neq 1$}
    In this case, we have two sets of critical points: $D_1=(0, 0, 0, s, v),\\ D_2=(0, 1, z, s, v)$ and two critical points: \\ $D_3=
    \Big(-\frac{\alpha\sqrt{6}(2\alpha w+2\alpha-w-3)}{2v_*(2\alpha-1)}, 0, 0,s_*, v_*\Big)$ and \\
    $D_4=\Bigg(-\frac{\sqrt{6}\alpha(\alpha-1)}{v_*(2\alpha-1)}, 0, \frac{3\Big(-\frac{3\alpha(\alpha-1)}{v_*(2\alpha-1)}\Big)^{-2\alpha}2^{-\alpha}}{2\alpha-1}, s_*, v_*\Bigg)$.  It can be seen from the second term of Eq. \eqref{xprime} that the existence of set $D_1$ demands $0<\alpha<\frac{1}{2}$ and $D_2$ is $0<\alpha<1$. Therefore, the set $D_1$ is not physically viable as it existence implies $C_s^2<0$.  Further, the existence of points $D_3$ and $D_4$ demands those functions $f$ for which $v_* \neq 0$.  In what follows, we discuss the stability conditions of each physically viable critical points.
    \begin{itemize}
   
        \item The set $D_2$ corresponds to an accelerated scalar field dominated solution $[\Omega_{m}=0$, $\Omega_\phi=1$, $w_{\rm eff}=-1, w_{\phi}=-1]$. It is non-hyperbolic in nature with eigenvalues $\Big\{\lambda_1=-3(1+w), \lambda_2=0, \lambda_3=0, \lambda_4=0, \lambda_5=0\Big\}$. The stability behaviour of points on this set can be determined numerically from the behaviour of perturbed trajectories near this set by taking a specific example of $f$ and $V$. It can be seen that trajectories near this set approach to different points on the set as $N \rightarrow \infty$ (see Fig. \ref{fig:pertt_d2}). Therefore, this set behaves as a late time attractor but not global as $y \neq 0$.
        
        \item The point $D_3$ corresponds to a matter dominated solution $[\Omega_{m}=1$, $\Omega_\phi=0$, $w_{\rm eff}=w, w_\phi=\frac{1}{2\alpha-1}]$. The eigenvalues of the corresponding perturbed matrix are \\ $\Big\{\lambda_1=\frac{3}{2}{\frac {2\,s_*\, \left( w+1 \right) {\alpha}^{2}+ \left( \left( 2\,w+2 \right) v_*-s_*\, \left( w+3 \right)  \right)  \alpha-v_*\, \left( w+1 \right) }{v_*\, \left( 2\,\alpha-1 \right) }}, \\  \lambda_2= \frac{3\alpha(2\alpha w +2\alpha-w-3)dg(s_*)}{v_*(2\alpha-1)},  \lambda_3= -\frac{3(2\alpha w +2\alpha-w-3)}{2(2\alpha-1)},\\ \lambda_4= \frac{3(2\alpha w -w-1)}{2\alpha-1}, \lambda_5= \frac{3\alpha(2\alpha w +2\alpha-w-3)dh(v_*)}{v_*(2\alpha-1)} \Big\}$.  Due to the complicated expressions of the eigenvalues on model parameters, we analysed its region of stability for a physically interesting case $w=0$. It can be seen that this point is either stable or saddle by determining the region of stability in $(\alpha,s_*,v_*)$ parameter space for the case $w=0$ (see Fig. \ref{fig:Region_Plot_D3_3D}). In fact the stability behaviour of this point is not cosmologically viable to describe the accelerated late time DE dominated universe, but its saddle behaviour is of cosmological interest as it corresponds to an intermediate  unaccelerated matter domination.

        \item The critical point $D_4$ also corresponds to a solution dominated by the kinetic energy part of the scalar field $[\Omega_m=0, \Omega_\phi=1, w_{\rm eff}=\frac{1}{2\alpha-1}, w_{\phi}=\frac{1}{2\alpha-1}]$. The eigenvalues of the corresponding perturbed matrix are $\Big\{\lambda_1=\frac{3 \alpha\left(s_*(\alpha-1)+v_*\right)}{v_*\left(2\alpha-1\right)},\\ \lambda_2= \frac{3 (1-\alpha)}{2\alpha-1}, \lambda_3=\frac{3\left(1-w(2\alpha-1)\right)}{2\alpha-1}, \lambda_4= \frac{6\alpha (\alpha-1)}{2\alpha-1} \frac{dg(s_*)}{v_*}, \lambda_5=\frac{6\alpha (\alpha-1)}{2\alpha-1} \frac{dh(v_*)}{v_*}\Big\}$. It can be either a stable point or saddle or unstable node depending on the choice of $\alpha$, $w$, the function $f$ and potential $V$. However, for the case $w=0$, this point is either saddle or unstable node only as one of the eigenvalue $\lambda_3$ is positive. In Fig. \ref{fig:RegionPlotD4}, we plot the region of instability of this point in $(\alpha,s_*,v_*)$ parameter space for the case $w=0$. 
        The saddle behaviour of this point is cosmologically rich as it shows the contribution of a non-canonical term to describe two important intermediate epochs in the history of the universe: the radiation dominated epoch for $\alpha=2$ ($w_{\rm eff}=\frac{1}{3}$) and the matter domination for very large values of $\alpha$ ($w_{\rm eff}=0$).
    \end{itemize}
    From the above analysis, we see that depending on the choice of function $f$ and the value of parameters $\alpha$, this class of non-canonical scalar field models can successfully describe the behaviour of the Universe at the background level. For $f$ in which $v_* \neq 0$ and $\alpha <1$, the models can describe the evolution of the Universe from a matter domination (point $D_3$) towards a deSitter attractor (set $D_2$). For $\alpha=2$, the models can explain the radiation domination (point $D_4$) to matter domination (point $D_3$) only, but these cannot be connected to a DE finite late time attractor.  Thus, in summary, the class of models in which $f$ is such that $v_* \neq 0$ can only describe either the transition of radiation $\rightarrow$ DM only without a DE epoch or DM $\rightarrow$ DE but without a radiation epoch. However, for $f$ in which $v_*=0$, the models cannot successfully describe the complete evolution of the Universe as there is no matter dominated critical point which is required to explain the large-scale structure formation at the background level.

    \subsubsection{$\alpha=1$}
    Here, we have two sets of critical points $E_1=(0, 0, z, s, v)$ and $E_2=(0, 1, z, 0, v)$. The set $E_1$ corresponds to a matter dominated solution $[\Omega_{m}=1$, $\Omega_\phi=0$, $w_{\rm eff}=w, w_\phi=\frac{1}{2\alpha-1}]$ with eigenvalues $\Big\{\lambda_1=\frac{3(w-1)}{2}, \lambda_2=\frac{3(1+w)}{2}, \lambda_3=0, \lambda_4=0, \lambda_5=0\Big\}$. Therefore, points on this set always behave as saddle points. The set $E_2$ corresponds to an accelerated scalar field dominated solution $[\Omega_{m}=0$, $\Omega_\phi=1$, $w_{\rm eff}=-1, w_\phi=-1]$. It is non-hyperbolic in nature with eigenvalues $\Big\{\lambda_1=0, \lambda_2=0, \lambda_3=-\frac{3}{2}\Big[1+\sqrt {1-\frac{4}{3}\frac{g(0)}{z}}\Big], \lambda_4=-\frac{3}{2}\Big[1-\sqrt {1-\frac{4}{3}\frac{g(0)}{z}}\Big],\\ \lambda_5=-3(1+w)\Big\}$. However, this set is normally hyperbolic set if $\frac{g(0)}{z}\neq 0$. Therefore, points on this set  behave as stable node for $0<\frac{g(0)}{z} \leq \frac{3}{4}$, they behave as stable focus for $\frac{g(0)}{z}>\frac{3}{4}$ and behave as saddle for $\frac{g(0)}{z}<0$. For the case  $\frac{g(0)}{z}= 0$, the set $E_2$ corresponds to a non-hyperbolic set. In order to confirm the stability behaviour of this set, we numerically perturbed the solution near the set and determine the behaviour of trajectories by considering a specific choice of $f$ and $V$. It is evident from Figs. \ref{fig:pert_e2_x}, \ref{fig:pert_e2_y} and \ref{fig:pert_e2_s} that the phase space trajectories approach to $x=0$, $y=1$ and $s=0$ respectively as $N \rightarrow \infty$. Further it can be seen from Fig. \ref{fig:pert_e2_z}, that trajectories starting from any values of $z$ remains almost constant. From the behaviour of the perturbed system near the set $E_2$, we can conclude that the set $E_2$ behaves as a stable set even for $\frac{g(0)}{z}=0$. Note that for the choice of potential considered in Fig. \ref{fig:pert_e2}, we have $g(0)=0$. From the above analysis, we see that in this case for some choice of potential and initial conditions, the universe passes through a matter domination epoch and evolve towards an accelerated DE dominated epoch. Hence, the model can successfully describe the late time transition of the universe.
    
    Finally, we end this section with some comments on the choice of coupling and potential functions. In the context of inflation, there are huge classes of potentials which give the same dynamics. Thus in order to connect observations with theories, it is important to identify possible degeneracy between two classes of potentials. For a detail discussion of degeneracy between non-canonical and canonical models in the context of inflation see \cite{gwyn,jarv}. Usually, the general $k$-essence Lagrangian of the form $\mathcal{L}(\phi, X)=f(\phi)X^\alpha-V(\phi)$ can always be recast  into a canonical kinetic term with a modified potential for $\alpha=1$ by a  field redefinition \cite{gwyn}. This is, however, non-trivial for a higher value of the exponent $\alpha$.  In general, by field redefinition, out of $f$ and $V$, one of them can be reduced to a trivial form for $\alpha=1$. This indicates that there may be some kind of correspondence between the results of $f=$ constant and $V=$ constant cases for $\alpha=1$. This is also suggested from the cosmological evolution of both the cases at $\alpha=1$, where the universe evolves from a stiff matter to a DM domination and then eventually evolves to a DE domination.

    \section{Conclusion}\label{conclusion}
    In the present work, we have performed a dynamical system analysis of a general $k$-essence  Lagrangian. The general $k$-essence models give interesting cosmological dynamics such as a viable cosmological transition from radiation to matter domination and eventually to DE domination. More importantly, we get a scenario in hand which gives the appealing unified description of DE and DM for a broad class of coupling functions.  The main aim of this work is to analyse the background cosmological dynamics for a broad class of coupling and potential functions. Dynamical system techniques allowed us to determine the broad class of coupling and potential functions which can lead to interesting cosmological dynamics.

    In the analysis presented above, we have first analysed the case when both the function $f$ and potential $V$ are constant  (see Sect. \ref{f_V_constant}). In this case, the system contains only a matter dominated solution  ($\Omega_m=1, w_{\rm eff}=w$) and a stable deSitter solution ($\Omega_\phi=1, w_{\rm eff}=-1$). Therefore, the dynamics is very simple and resembles that of the $\Lambda$CDM where the Universe evolves from matter domination towards a DE domination. Secondly, we have considered the case when function $f$ is still constant but the potential is not constant (Sect. \ref{f_constant}). In both the cases, for all physically viable critical sets (or points), we have $x=0$ which implies that there is no contribution from the kinetic component of the scalar field during the main cosmological eras. 
    
    Then we have analysed the case when the function $f$ is assumed to be such that $\Gamma_2$ is a function of $v$. Here, the DM contribution comes from the extra barotropic fluid alone except for the case when the function $f$ is such that $v_*\neq 0$ ($v_*$ is a root of $v^2(\Gamma_2(v)-1)=0$) along with a constant potential. In the latter case, the early cosmic matter behaviour is mainly due to the kinetic part of the non-canonical scalar field (point $C_5$ in Sect. \ref{v_constant}).  Depending on the value of the non-canonical parameter $\alpha$,  the Universe evolves either from a radiation or DM or stiff matter solution described by the point $C_5$  towards a DM dominated epoch and eventually settles as a cosmological constant (see Fig. \ref{fig:Cos_Par2} in Sect. \ref{v_constant}). In particular for $\alpha=2$, the model provides a viable cosmological sequence: radiation [$w_{\rm eff}=\frac{1}{3}$] $\rightarrow$ DM [$w_{\rm eff}=0$] $\rightarrow$ DE [$w_{\rm eff}=-1$]. This result extends the expansion history of the Universe obtained in \cite{Sahni:2015hbf}. Note that the early radiation behaviour is solely due to the contribution of the scalar field without a need for normal radiation. Such interesting dynamics cannot be achieved in the case of a canonical scalar field within the GR framework which indeed requires additional normal radiation for identical dynamics.  It is worth mentioning, the identical sequence of cosmological background dynamics can also be obtained in the context of mimetic gravity \cite{Dutta:2017fjw}. Moreover, as the non-canonical parameter $\alpha$ affects the dark energy speed of sound, therefore depending on the clustering property of a scalar field, the early non-canonical scalar field domination behaves either as cold or warm dark matter \cite{Sahni:2015hbf}. This might somehow circumvent the problem of $\Lambda$CDM in fitting CMB and weak lensing data simultaneously \cite{Kunz:2015oqa}. Further, the early dark matter behaviour of the kinetic component of a scalar field could also explain the formation of seeds for super-massive black holes \cite{Sawicki:2013wja}. Interestingly, the scenario in hand can also explain a possible inflationary exit problem (point $C_4$) for a constant potential. This behaviour is also achieved in $k$-essence model discussed in \cite{DeSantiago:2012nk}. To have a clear picture on this, one requires to investigate the dynamical behaviour of the potential $V$ and determine the conditions for evolving towards a slow-roll regime. It deserves mentioning that a similar form of a solution is however also obtained in a well-known $\alpha$-attractors inflationary models with non-constant potential \cite{Alho:2017opd}. Here, we should remark that in $\alpha$-attractors inflationary models, the slow roll conditions are usually satisfied when the coupling function $f$ is very large, irrespective of the form of the potential $V$. In particular, a singular $f$ may correspond to a critical point, which however cannot be captured by variables \eqref{dmv}.  The critical point for such behaviour might be pushed towards infinity and compactification techniques may give such a point. Further, it is worth noting that due to the absence of accelerated global attractors, the late time evolution from decelerated to an accelerated phase is not guaranteed but strictly depends on fine-tuning of initial conditions and model parameters.  In addition to this, the non-compact form of the phase space demands the compactification technique of the phase space, so as to have a global picture on the cosmological dynamics. However, such an analytical investigation is beyond the scope of the present work.
    
 It is worth noting that in the presence of the photons, the radiation dominated era would comprise of two components i.e. the scalar field kinetic component and photons that cause the space to expand. However, only one component i.e. photons that are emitted and absorbed by matter. Despite  tight constraints from standard big bang nucleosynthesis, this may signal interesting observational characteristics and perhaps could be connected with the puzzle of the 21 cm line anomaly. The discussion on the implications of this phenomenon will require further investigations  in the future.
    
    In summary, we conclude that on employing dynamical system techniques, the non-canonical scalar field models give contrasting dynamics for different choices of $f$ and $V$.  More precisely, our approach identified a broad class of coupling and potential functions which give interesting cosmological dynamics at the background level. In particular, our analysis reveals that a class of coupling functions $f$ where the root of an equation $v^2(\Gamma_2(v)-1)=0$ does not vanish and with a constant potential  lead to interesting cosmological dynamics viz. a viable history of the Universe (radiation $\rightarrow$ DM $\rightarrow$ DE dominated era) as well as the unified description of DM and DE.  It would be of interest in the future to further analyse the behaviour using an alternative choice of dynamical variables which might give a more broad class of $f$ and $V$ to describe the dark sectors unification. This would corroborate the work of  \cite{Mishra:2018tki}. Further, for a various choice of $f$ and $V$, this model can mimic the $\Lambda$CDM model in late times at the background level which supports the work of Sahni and Sen \cite{Sahni:2015hbf}. However, there may be a possible deviation at the perturbation level, which might lead to interesting signatures towards the present and future observations \cite{Sawicki:2013wja}. By employing dynamical system techniques, analysis at the perturbation level might give a general conclusion over a wide range of initial conditions for perturbations. Finally, in the light of precise cosmological data, it would be of interest to investigate the degree of degeneracy of this class of models for higher order kinetic terms. 
    
    \acknowledgement
    
    The authors thank Laur J\"arv for useful discussions. JD was supported by the Core research grant of  SERB, Department of Science and Technology India (File No.CRG/2018/001035)  and  the Associate program of IUCAA.
     The authors also thank the anonymous reviewer for  constructive suggestions which lead
to the improvement of the work.

\end{document}